\documentclass[final,5p,times,twocolumn]{elsarticle}

\usepackage{amssymb}
\usepackage{multirow}

\biboptions{square,sort&compress}
 
 \def\NIMA{{Nucl. Instrum. Methods} A}
\def\Journal#1#2#3#4{{\em#1} {\bf #2}, #3 (#4)}
\def\mr#1{\multirow{2}{*}{#1}}

\usepackage{lineno}

\usepackage[german,english]{babel}
\usepackage{subfigure}
\usepackage{epsfig}
\usepackage{upgreek}
\usepackage{amsmath}
\usepackage{draftcopy}
\usepackage{xfrac}
\usepackage{xcolor} 
\usepackage{color,soul} 

\begin{document}

\begin{frontmatter}



  \title{Thermo-acoustic Sound Generation in the Interaction of Pulsed
    Proton and Laser Beams with a Water Target}

\author[ecap]{R.\ Lahmann\corref{cor1}}\ead{robert.lahmann@physik.uni-erlangen.de}
\author[ecap]{G.\ Anton}
\author[ecap]{K.\ Graf}
\author[ecap]{J.\ H{\"o}{\ss}l}
\author[ecap]{A.\ Kappes}
\author[ecap]{U.\ Katz}
\author[theo]{K.\ Mecke}
\author[heidelberg]{S.\ Schwemmer\fnref{ref1}}

\address[ecap]{Erlangen Centre for Astroparticle Physics (ECAP),
  Friedrich-Alexander-Universit\"at Erlangen-N\"urnberg,  
  Erwin-Rommel-Str.\ 1, 91058
  Erlangen, Germany} 

\address[theo]{Institut f\"ur Theoretische Physik, 
Friedrich-Alexander-Universit\"at Erlangen-N\"urnberg,  
Staudtstr.\ 7, 91058 Erlangen, Germany}

\address[heidelberg]{Landessternwarte, K{\"o}nigstuhl 12, 69117 Heidelberg, Germany (formerly at ECAP)}

\cortext[cor1]{Corresponding author}
\fntext[ref1]{formerly at ECAP}
\address{}

\begin{abstract}
  The generation of hydrodynamic radiation in interactions of
  pulsed proton and laser beams with matter is explored.  The beams were directed
  into a water target and the resulting acoustic signals were recorded
  with pressure sensitive sensors.  Measurements were performed with
  varying pulse energies, sensor positions, beam diameters and
  temperatures.  The obtained data are matched by simulation results
  based on the thermo-acoustic model with uncertainties at a level of 10$\%$.  
  The results imply that the primary mechanism for sound
  generation by the energy deposition of particles propagating in
  water is the local heating of the medium. The heating results in a
  fast expansion or contraction and a pressure pulse of bipolar
  shape is emitted into the surrounding medium. An interesting, widely discussed
  application of this effect could be the detection of ultra-high
  energetic cosmic neutrinos in future large-scale acoustic neutrino
  detectors. For this application a validation of the sound generation
  mechanism to high accuracy, as achieved with the 
  experiments discussed in this article, is of high importance.
\end{abstract}

\begin{keyword}
  cosmic neutrinos \sep acoustic neutrino detection \sep thermo-acoustic model \sep ultra-high
  energy cosmic rays \sep beam interaction \PACS



\end{keyword}

\end{frontmatter}


\section{Introduction}

In 1957 G.A.\ Askaryan pointed out that ionisation and cavitation
along a track of an ionising particle through a liquid leads to
hydrodynamic radiation \cite{Askaryan1}.  In the 1960s, 1970s and
1980s, theoretical and experimental studies have been performed on the
hydrodynamic radiation of beams and particles traversing dense media
\cite{Askaryan2, Learned, Sulak, Hunter1, Hunter2, Albul1, Albul2}.

The interest in characterising the properties of the acoustic
radiation was, among other reasons, lead by the idea that the effect
can be utilised to detect ultra-high energy ($E \gtrsim
10^{18}\,\textrm{eV}$) cosmic, i.e.\ astrophysical neutrinos, in dense
media like water, ice and salt.  In the 1970s this idea was discussed
within the DUMAND optical neutrino detector project \cite{Dumand} and
has been studied in connection with Cherenkov neutrino detector projects since. 
The detection of such neutrinos is considerably more challenging
than the search for high-energy neutrinos ($E \gtrsim
10^{10}\,\textrm{eV}$) as currently pursued by under-ice and
under-water Cherenkov neutrino telescopes~\cite{icecube,antares,baikal}. Due to the low expected
fluxes, detector sizes exceeding 100\,km$^3$ are needed \cite{seckel}. 
However, the properties of the
acoustic method allow for sparsely instrumented arrays with $\sim$100
sensors/km$^3$.

To study the feasibility of a detection method based on acoustic
signals it is necessary to understand the properties of the sound
generation by comparing measurements and simulations based on theoretical
models. According to the so-called thermo-acoustic model
\cite{Askaryan2}, the energy deposition of particles traversing
liquids leads to a local heating of the medium which can be regarded
as instantaneous with respect to the hydrodynamic time scales. Due to
the temperature change the medium expands or contracts according to
its bulk volume expansion coefficient $\alpha$. The accelerated motion of
the heated medium generates an ultrasonic pulse whose temporal signature
is bipolar and which propagates in the volume. 
Coherent superposition of the elementary sound waves, produced over the cylindrical volume of the energy deposition, 
leads to a propagation within a flat disk-like volume in the direction perpendicular to the axis of the particle shower.

In this study, the hydrodynamic signal generation by two types of
beams, interacting with a water target, was investigated: pulsed protons and a pulsed laser, 
mimicking the formation of a hadronic cascade from a neutrino interaction under laboratory conditions. 
With respect to the aforementioned experimental studies of the thermo-acoustic model, the work presented here can make use of previously unavailable advanced tools such as GEANT4%
~\cite{geant4}
for the simulation of proton-induced hadronic showers in water. 
Good agreement was found in the comparison of the measured signal
properties with the simulation results,
providing confidence to apply similar simulation methods
in the context of acoustic detection of ultra-high energy neutrinos.
A puzzling feature observed in previous studies---a non-vanishing signal amplitude 
at a temperature of 4$^\circ$C, where for water at its highest density
no thermo-acoustic signal should be present---was investigated in detail.
Such a residual signal was also observed for the proton beam experiment described in this article, but not for the laser beam, indicating that the formation of this signal is related to the charge or the mass of the protons.

\section{Derivation of the Model}\label{sec_model} 

In the following, the thermo-acoustic model \cite{Askaryan2, Learned}
is derived from basic assumptions, using a hydrodynamic approach.
Basis is the momentum conservation, i.e., the Euler Equation
\begin{equation}
\frac{\partial (\rho v_i)}{\partial t} = - \sum_{j=1}^3 \frac{\partial \Pi_{ij}}{\partial x_j}\textrm{\ ,}
\label{eq_euler}
\end{equation}
for mass density $\rho$, velocity vector field of the medium $\left(v_1, v_2, v_3\right)$\  and momentum-density tensor
\begin{equation}
 \Pi_{ij} = p\delta_{ij} + \rho v_i v_j
\label{eq_densitytensor}
\end{equation}
including the pressure $p$~\cite{Landau}. Equation (\ref{eq_euler}) can be derived
from momentum conservation. 
In the derivation, energy dissipation resulting from processes such as internal friction or heat transfer
are neglected. Motions described by the Euler Equation hence are adiabatic. 
Taking the three partial derivatives of Eq.\ (\ref{eq_euler}) with respect to $x_i$ 
 and using for the density the continuity equation
\begin{equation}
\frac{\partial \rho}{\partial t} + \sum_{i=1}^3 \frac{\partial }{\partial x_i} (\rho v_i)= 0\textrm{\ ,}
\label{eq_continuity}
\end{equation}
a non-linear wave equation can be derived:
\begin{equation}
\frac{\partial^2 \rho}{\partial t^2}= \sum_{i,j=1}^3 \frac{\partial^2 \Pi_{ij}}{\partial x_i \partial x_j}\,.
\label{eq_wave_base}
\end{equation}

To solve this equation, the problem is
approached in two separated spatial regions: Firstly, a region $B$
(\textit{`beam'}), where the energy is deposited in the beam
interactions with the fluid and thus the wave excited in a non-equilibrium process; and secondly, a
hydrodynamic (\textit{`acoustic'}) region $A$, where
the acoustic wave propagates through the medium and where linear hydrodynamics in local equilibrium can be assumed.  
This splitting can
be reflected by the momentum density tensor, rewriting it as
\begin{equation}
\Pi_{ij}(\vec{r})= \left\{\begin{matrix}\Pi_{ij}^A(\vec{r}) & \text{for } \vec{r}\in A\cr
 \Pi_{ij}^B(\vec{r}) & \text{for } \vec{r}\in B \end{matrix}\right. \textrm{\ .}
\end{equation}
In local equilibrium the changes in mass density are given by 
\begin{alignat}{2}
  d\rho &\,  = &\, \frac{\partial \rho}{\partial p}\Bigg|_{S,N} dp & + \frac{\partial \rho}{\partial S}\Bigg|_{p,N}dS \cr 
  &\, = &\,  \frac{1}{v_s^2}dp  &- \frac{\alpha}{c_p}\frac{\delta Q}{V}  
  \label{eq_density_change}
\end{alignat}
with the bulk volume expansion coefficient
$\alpha=-{\frac{1}{\rho}}{\frac{\partial \rho}{\partial T}}|_{p,N}$, the 
energy deposition  $\delta Q = TdS$, the adiabatic speed of sound $v_s$ in the medium and the specific heat $c_p={T\over N}{\partial S\over\partial T}|_{p,N}$. 

In the acoustic regime, where $\delta Q = 0$, the momentum density tensor can be expressed as
$\Pi_{ij}^A= p\delta_{ij} =  v_s^2\rho\delta_{ij}$ (using Eqs.\
(\ref{eq_densitytensor}) and (\ref{eq_density_change})), where we assume an adiabatic density change with pressure. 
The non-linear kinetic term $ \rho v_i v_j$ entering $\Pi_{ij}^A$ according to Eq.~(\ref{eq_densitytensor}) can be neglected 
for small deviations $\rho' = \rho - \rho_0$ from the static density $\rho_0$ and small pressure differences $p' = p - p_0$ from the static pressure $p_0$ \cite{Landau}.

In the region B, where non-equilibrium deposition occur, one may make the {\it ansatz} 
\begin{equation}  
\Pi_{ij}^B = p\delta_{ij} + \beta u_iu_j
\label{eq:ansatz}
\end{equation} 
with the direction $u_i$ of the beam which breaks the isotropy of the energy-momentum tensor and describes with the parameter $\beta$ in an effective way the momentum transfer on the fluid. 

Although in non-equilibirum we apply Eq.~(\ref{eq_density_change}) with the energy deposition density $\epsilon\equiv Q/V$ of the beam. Then, 
with the additional energy-momentum tensor due to the beam 
$$
\delta \Pi_{ij}^B := v_s^2\frac{\alpha}{c_p}{{\epsilon}}\delta_{ij} +\beta u_iu_j 
$$ 
the wave equation (\ref{eq_wave_base}) reads
\begin{equation}
\vec{\nabla}^2 p' - \frac{1}{v_s^2}\frac{\partial^2 p'}{\partial t^2} = 
 -\sum_{i,j=1}^3 \frac{\partial^2 \delta\Pi_{ij}^B}{\partial x_i \partial x_j}
\end{equation}
The general solution for the wave equation  can be
written using a Green function approach as 
\begin{eqnarray}
p'(\vec{r},t) &= & {1 \over 4\pi }  \sum_{i,j=1}^3 \int_B dV' \frac{1}{|\vec{r}-\vec{r}\,'|} {\partial^2 \delta\Pi^{B}_{ik}(\vec{r}',t') \over \partial x'_i \partial x'_k} \cr
&= &   {1 \over 4\pi  }  \int_B dV' \left[
{n_in_k 
\over |\vec{r}-\vec{r}\,'| } \frac{\partial^2 \delta\Pi^{B}_{ik}(\vec{r}',t')}{v_s^2 \partial {t'}^2} \right.   
\label{eq:pressure-approx}
\\
 & +  &  {3n_in_k-\delta_{ik}\over |\vec{r}-\vec{r}\,'|^2 }\left. \left(\frac{\partial \delta\Pi^{B}_{ik}(\vec{r}',t')}{v_s \partial t'}  
+ \frac{\delta\Pi^{B}_{ik}}{|\vec{r}-\vec{r}\,'|} \right)  \right]
\nonumber 
\end{eqnarray}
with the components of the unit vector $n_i = (x_i - x'_i)/|\vec{r} - \vec{r}\,'|$  and the retarded time $t'=t-|\vec{r}-\vec{r}\,'|/v_s$.  
For the last conversion, partial integration and the total
derivative $\frac{d}{dx_i}=\frac{\partial}{\partial x_i}+\frac{1}{v_s}\frac{\partial}{\partial t}$ have been used repeatedly.  
Note that $\delta\Pi_{ij}^B  =0$ for $\vec{r}\in A$, so that the integration is carried out over the volume of the energy deposition region $B$.

Assuming an energy deposition without momentum transfer to
the medium, the kinetic term in the {\it ansatz} (\ref{eq:ansatz}) can be neglected ($\beta=0$) yielding  
\begin{equation}
  p'(\vec{r},t) = \frac{1}{4\pi}\frac{\alpha}{c_p}\int_V \frac{\mathrm{d}V'}{|\vec{r} - \vec{r}\,'|}\;\frac{\partial^2}{\partial t^2}\,\,\epsilon\left(\vec{r}\,',t'\right) 
\label{eq_pressure}
\end{equation}
for a thermo-acoustic wave generated solely by heating of the medium. 
The signal amplitude $p'$ can be shown to be proportional to the dimensionless quantity $v_s^2\, \alpha/c_p$
when solving Eq.~(\ref{eq_pressure}) 
for the case of an instantaneous energy deposition.

Equation~(\ref{eq_pressure})
is equivalent to the results obtained from the approaches presented in 
\cite{Askaryan2, Learned}. 
The derivation pursued above, however, uses a 
different approach starting with the Euler equation and an anisotropic energy-momentum tensor,
yielding a more general expression in Eq.~(\ref{eq:pressure-approx}).
Only if assuming an isotropic energy deposition one arrives at the expression for the pressure deviation $p'$ 
given in Eq.~(\ref{eq_pressure}). 
 
Note that the validation of the last assumption $\beta=0$ being a good approximation would require a detailed knowledge of the momentum transfer from the beam to the medium. Taking it into account would result in an additional dipol term 
$\sim \ddot{\beta}\,(\vec{u}\cdot\vec{n}\,)^2$ in Eq.~(\ref{eq_pressure}) which may become the dominant contribution to wave generation if $\alpha\approx 0$ close to 4.0$^{\circ}$C.  
However, for $\beta=0$ the pressure field resulting from a beam interaction in a medium is determined by the spatial and temporal distribution of the energy
deposition density $\epsilon$ alone. The amplitude of the resulting acoustic
wave is governed by the thermodynamic properties $v_s$, $c_p$ and
$\alpha$, the latter three depending primarily on the temperature of
the medium. A controlled variation of these parameters in the
conducted laboratory experiments and a study of the resulting pressure
signals therefore allows for a precise test of the thermo-acoustic
model.

Simulations based on the thermo-acoustic model, as performed to interpret the results of the experiments described in the next section, will be discussed in Sec.~\ref{sec_simulation}. 
Note that the energy deposition density $\epsilon$ and its temporal evolution for the proton and laser beam interactions discussed in this paper are quite different from those expected for the interaction of ultra-high energy neutrinos. However, if a simulation starting from basic principles allows for a good reproduction of the experimental results, it is reasonable to assume 
that these simulation methods are transferable to neutrino interactions, as they are governed by the same underlying physical processes.

\section{Experimental Setup and Beam Characteristics}\label{sec_setup}

\begin{figure}[bht]
  \begin{center}
    \subfigure[Top view]{\parbox[c]{7.5cm}
      {\includegraphics[width=7.5cm]{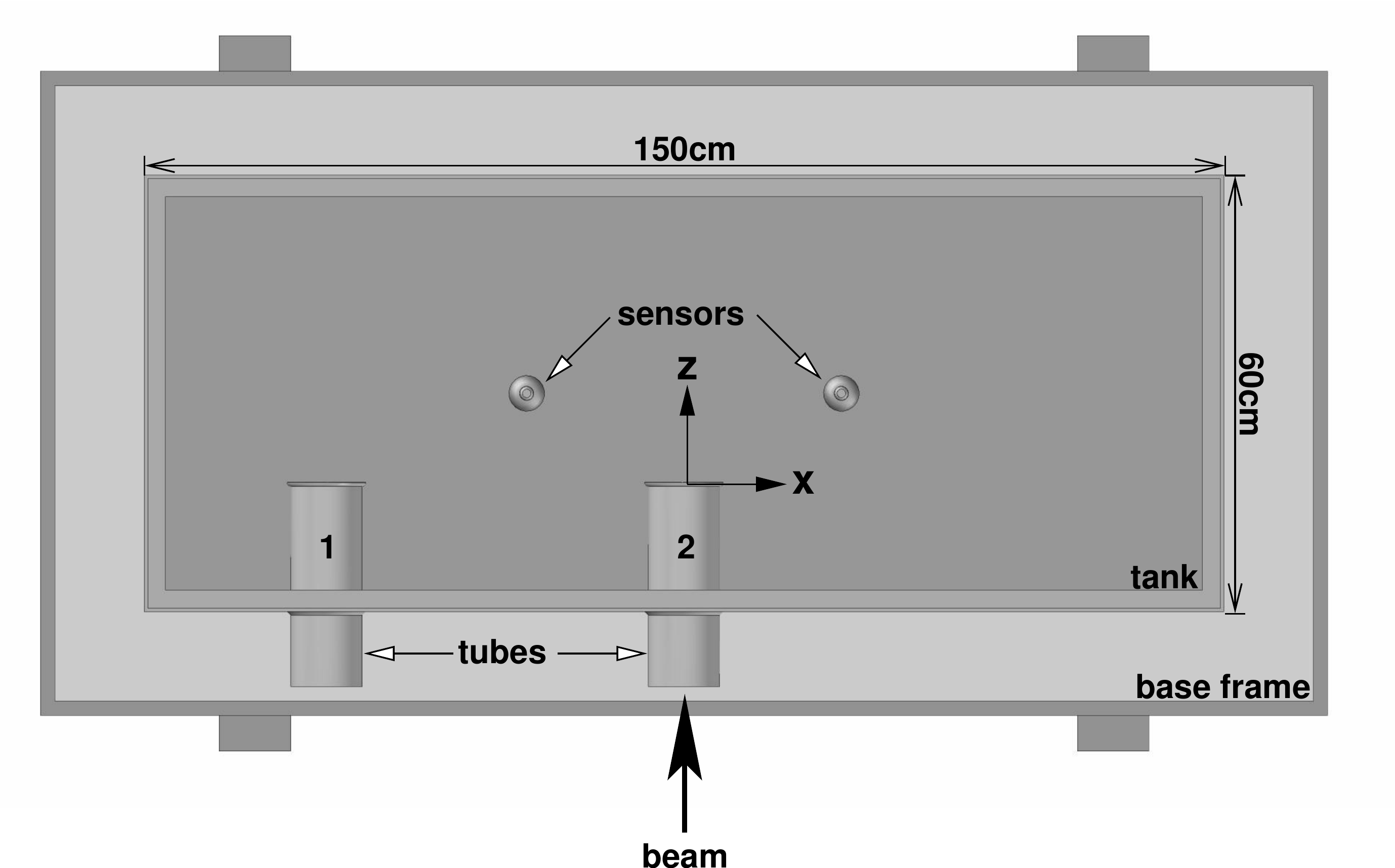}\label{fig_test_setup_a}}}
    \hspace{\subfigtopskip}
    \subfigure[Isometric view]{\parbox[c]{4.9cm}
      {\includegraphics[width=4.9cm]{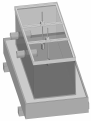}\label{fig_test_setup_b}}}
    \caption{Schematic view of the experimental setup. Beams can 
      enter the water volume through a thin foil (proton beam) or glass
      window (laser beam), which are set away from the tank walls via
      an air-filled tube to minimise effects of surface reflections 
      on the recorded signals. 
 Either of the two tubes, labelled `1' and `2' in \subref{fig_test_setup_a} was used, depending on the experimental
 setup.
      The origin of the coordinate system was chosen at the 
      entry point of the beam into the water target, see example in \subref{fig_test_setup_a}.     
	\label{fig_test_setup} }
  \end{center}
\end{figure}

The experiments presented in this paper were performed with a pulsed
infrared Nd:YAG laser facility ($\lambda = 1064 \, \mathrm{nm}$)
located at the Erlangen Centre for Astroparticle Physics (ECAP) 
of the University of Erlangen, and the pulsed $177 \,
\mathrm{MeV}$ proton beam of the ``Gustaf Werner Cyclotron'' at the
``Theodor Svedberg Laboratory'' in Uppsala, Sweden.  The beam
properties allow for a compact experimental setup. In both cases, the
beams were dumped into a dedicated $150 \times 60 \times 60 \,
\mathrm{cm}^3$ water tank, where the acoustic field was measured with
several position-adjustable acoustic sensors (see
Fig.~\ref{fig_test_setup}). The sensors (also called hydrophones) could be positioned within the
tank with absolute uncertainties below 1\,cm.  The temperature of the
water could be varied between $1^\circ \mathrm{C}$ and $20^\circ
\mathrm{C}$ with a precision of $0.1^\circ \mathrm{C}$. 
The temperature was brought to a particular value by first
cooling the water with ice; subsequently the whole water volume was
heated to the desired temperature in a controlled, gradual procedure.
Once the water temperature had been established, at least 10\,min
remained for measurements until the water volume heated up by
$0.1^\circ \mathrm{C}$ through heat transfer from the environment.
This time span was sufficient for all measurements conducted at 
water temperatures below the ambient temperature.

\begin{figure}[htb]
  \begin{center}
    \includegraphics[width=7.8cm]{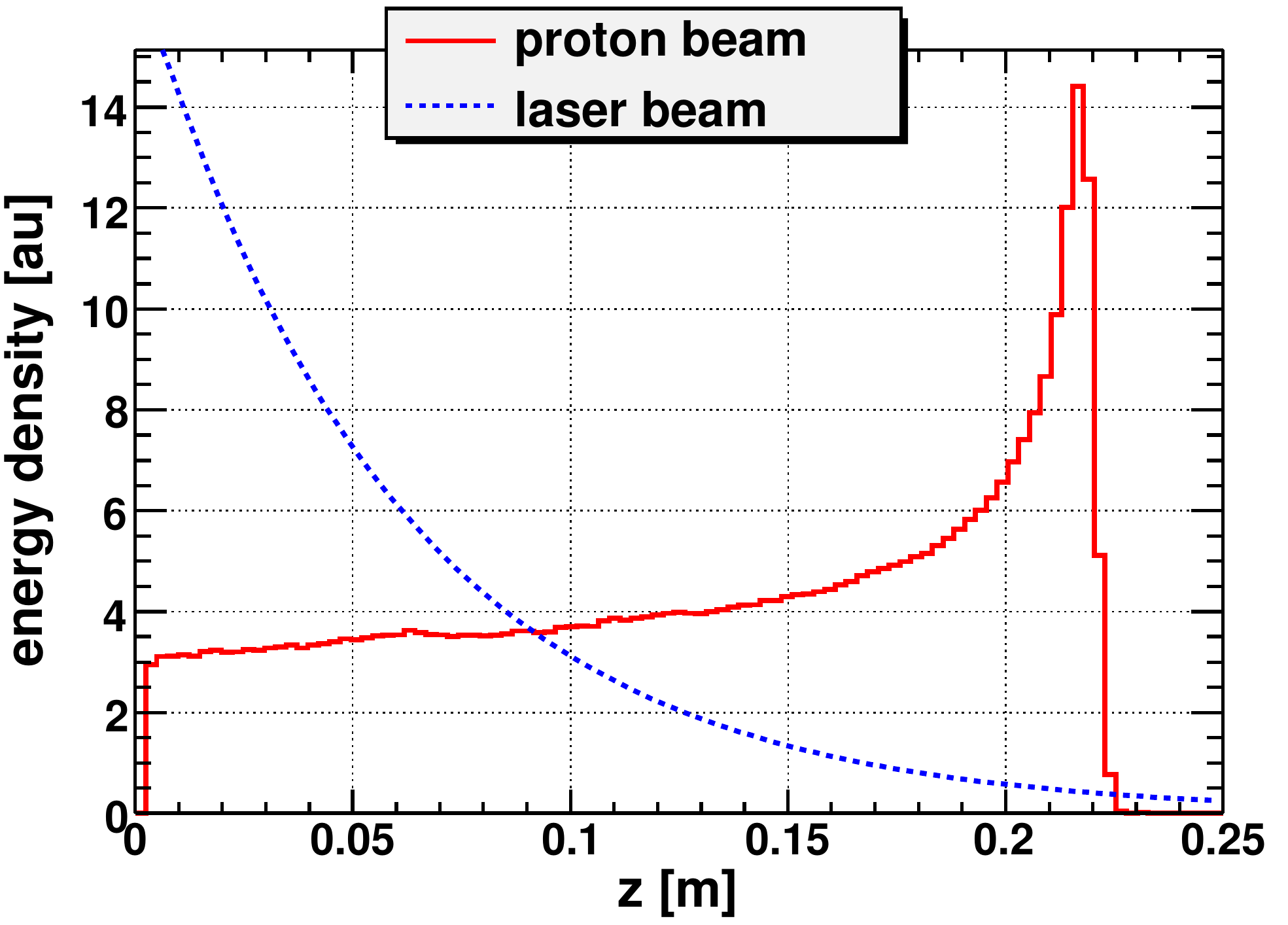}
    \caption{Simulated energy deposition density along the beam axis (z-axis)
      for the two beam types (integral normalised to 1). For the
      proton beam the Bragg peak is prominent at 0.22\,m, for the 
      laser beam the decrease is exponential with an attenuation 
      length of 0.059\,m.
      \label{fig_energy_deposition_z}
    }
  \end{center}
\end{figure}

The explored range of spill energies for the proton beam was from $10
\, \mathrm{PeV}$ to $410 \, \mathrm{PeV}$, the beam diameter was
approximately $3 \, \mathrm{mm}$ and the spill time $t_s \approx 30 \,
\mathrm{\upmu s}$.  For $177\,\mathrm{MeV}$ protons, the energy
deposition in the water along the beam axis ($z$-axis, beam entry into
the water at $z=0\,\mathrm{cm}$) is relatively uniform up to
$z=20\,\mathrm{cm}$ ending in the prominent Bragg-peak at $z
\approx22\,\mathrm{cm}$ (see Fig.~\ref{fig_energy_deposition_z}). 
To adjust the spill energy, the number of protons per bunch was varied. The total charge of a bunch was calibrated with two independent methods (Faraday cups and scintillation counters), leading to an uncertainty on the order of 
15\%, with some higher values for low spill energies.
To obtain the beam intensity and profile for the proton interactions in the water tank,
the distance of about 1.2\,m that the beam was travelling from the exit of the beam pipe through air and its entering into the water tank were included
in the GEANT4 simulation.

For the laser experiment, the pulse energy  
was adjusted between $0.1 \, \mathrm{EeV}$ and $10\, \mathrm{EeV}$ and calibrated 
using a commercial power meter. The beam had a diameter of approximately $2$\,mm and the
pulse length was fixed at $9 \, \mathrm{ns}$. For the infrared light
used, the laser energy density deposited along the beam axis has an
exponential decrease with an absorption length of $(5.9\pm0.1) \,
\mathrm{cm}$ (see Fig.~\ref{fig_energy_deposition_z}). 

For both beam types the
lateral energy deposition profile was Gaussian (the
aforementioned beam diameters are the $\sigma$'s of the profiles). 
The two experiments allow to explore different spatial and
temporal distributions of the energy deposition as well as two
different mechanisms of energy transfer into the medium. For both
beams, energy is deposited via excitation, in addition the medium is
ionised in the case of the proton beam.

For the signal recording, sensors based on the piezo-electric effect
\cite{Anton} were used. A full characterisation of these sensors had
been performed prior to the experiments. They are linear in amplitude
response, the frequency response is flat starting from a few kHz up to
the main resonance at $40\,\mathrm{kHz}$ with a sensitivity of
$-155\,\mathrm{dB\,re\,}1\mathrm{V}/\upmu\mathrm{Pa}$ 
($\sim$0.02\,V/Pa).  The main resonance is more sensitive by 
$\sim$5$\,\mathrm{dB\,re\,}1\mathrm{V}/\upmu\mathrm{Pa}$ and sensitivity
drops rapidly at higher frequencies; 
at 90\,kHz, the sensitivity has dropped by 20\,dB.
The absolute uncertainty in the
determination of the sensitivity is at a level of 2\,dB in the frequency range of interest.
Above 90\,kHz the uncertainty exceeds 5\,dB. 
To calculate the response of the
sensors to an external pressure pulse a parametrised fit of an
equivalent circuit model as described in \cite{Anton} was used.  The
sensitivity dependence on temperature was measured and the relative
decrease was found to be less than $1.5\%$ (or about 0.13\,dB) per $1^{\circ}\mathrm{C}$.

For every set of fixed experimental parameters (temperature, energy,
sensor position, etc.)  the signals of 1000 beam pulses were recorded
with a digital oscilloscope
at a sampling rate in excess of $1\,\mathrm{MHz}$. This rate is sufficient for
the signals with spectral components up to 100\,kHz, where the
sensitivity of the sensors is negligibly small. These individual
pulses were averaged to reduce background and environmental noise in
the analysis, thereby obtaining a very high statistical precision.

\section{Basic Features of the Measured Signals}

\begin{figure}[ht]
  \begin{center}
    \includegraphics[width=7cm]{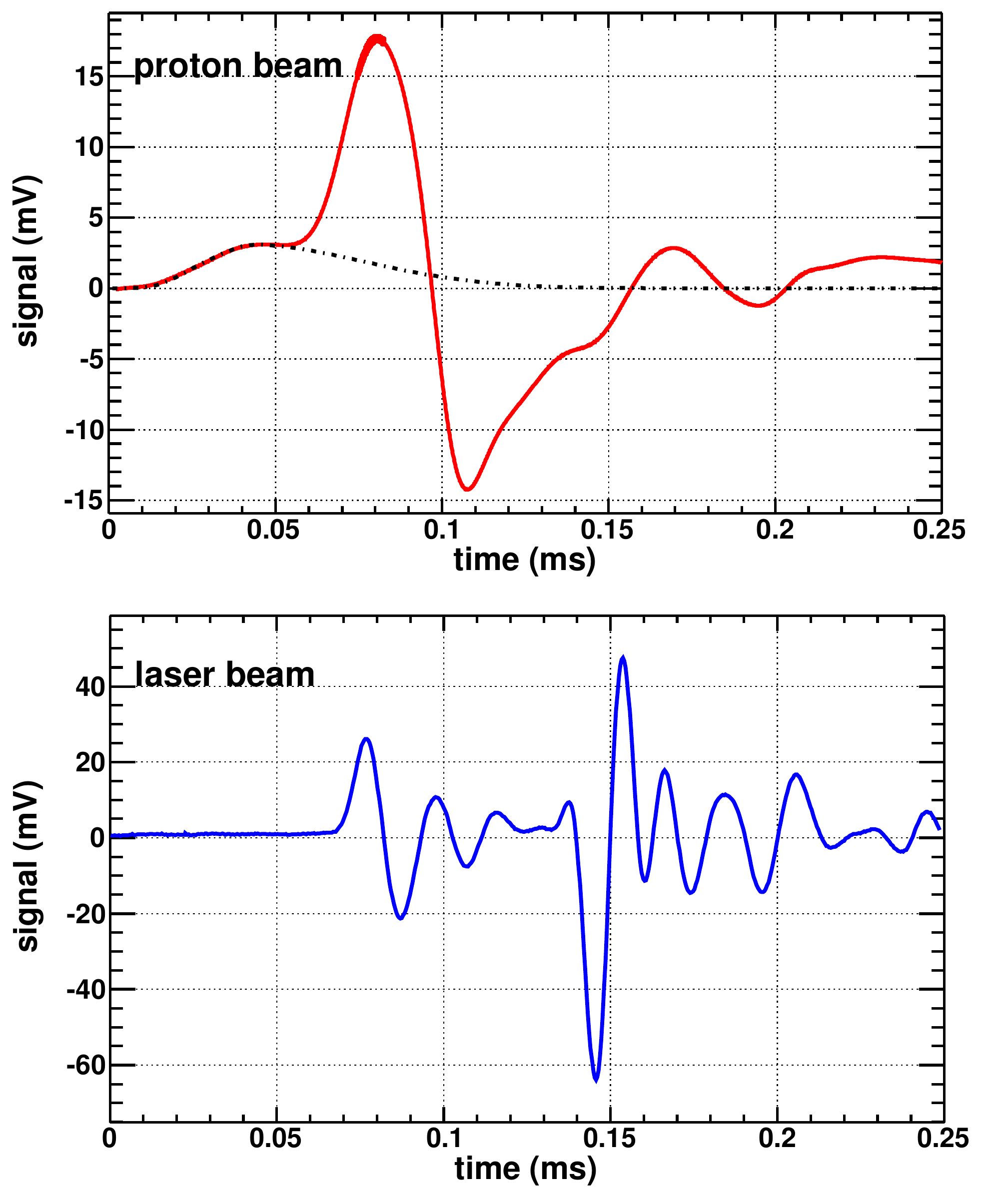}
    \caption{Typical signals measured in the proton (upper graph) and
      laser (lower graph) experiment. Both signals (solid lines) were
      taken at a sensor position near $x$=10\,cm and $z$=20\,cm and
      are shown within the same time interval. The dash-dotted line in
      the upper graph indicates the charge effect described in the
      text.
      \label{fig_signals} }
  \end{center}
\end{figure}

Figure \ref{fig_signals} shows typical signals measured in the proton
and the laser experiment, respectively, using the same sensors and experimental
setup.  The general shapes of the two signals differ: a typical signal
for the proton beam shows a bipolar signature\footnote{Note that also
  the expected signal shape for acoustic pulses produced in
  interactions of ultra-high energy neutrinos in water is a bipolar
  one \cite{Bevan}.}, the one for the laser deviates from such a
generic form. The laser signal has high frequency components up to
several MHz due to the high energy deposition density at the point of
beam entry and the almost instantaneous energy deposition compared to
the $\upmu$s pulse of the proton beam; therefore the resonance of the
sensor is excited causing a ringing in the measured signal. The
spatial distribution of the energy density $\epsilon$ deposited by the laser leads to the 
two separate signals:
the first originates in the beam area at the same $z$-region as the
sensor placement ({\it `direct signal'}), the second from the beam
entry, a point of discontinuity where most of the energy is deposed
({\it `beam entry signal'}).

The signal of the proton beam is deteriorated with respect to an ideal
bipolar signal. 
Three main contributions to this distortion can be discerned:
the recorded signal starts
before the expected onset of the acoustic signal (55.2\,$\upmu$s for
the given position, see Fig.~\ref{fig_signals}, 
given by the sonic path length); reflections of
the acoustic wave on the beam entry window overlay the original wave
starting in the first rare-faction peak; and finally there are
frequency components of the signal exciting a resonant response of the
sensor, 
slightly changing the signal shape and causing ringing.  The
first point was studied and found to be consistent with an electric
charge effect in the sensors caused by the proton beam. 
Its starting time was always coincident with 
the beam pulse entry into the water, even for
sensor distances of up to 1\,m, hinting at an electromagnetic
origin of the distorting signal.
The shape of this non-acoustic signal
is consistent with the integrated time-profile of
the beam pulse with a subsequent exponential decay.  This deformation
of the main signal is considered a systematic uncertainty on the signal
properties and treated as such in the analysis. For the most part, 
its shape was fitted and subtracted from the signal.

In order to minimise the impact of the signal deformation caused by the
described effects on the analysis of the recorded signals, robust
characteristics were used: the peak-to-peak amplitude and the signal
length from maximum to minimum of the signal. For the laser experiment
these features were extracted for the direct signals only.

\section{Simulation of Thermo-Acoustic Signals}\label{sec_simulation}

For an in-depth validation of the thermo-acoustic model, comparisons
of the signal properties with simulation results based on the model
are essential. To this end, a simulation of the expected signals was
developed. It is based on the thermo-acoustic model using a numeric
solution of Eq.\ (\ref{eq_pressure}). The input parameters to the
simulation were either measured at the experiments, i.e.\ medium
temperature and beam profiles, or simulated, i.e.\ the energy
deposition of the protons (using GEANT4). The thermodynamic parameters
bulk volume expansion coefficient, heat capacity and speed of sound  were
derived from the measured water temperature using standard
parametrisations. Tap water quality was assumed.

\begin{figure}[ht]
  \begin{center}
    \includegraphics[width=7cm]{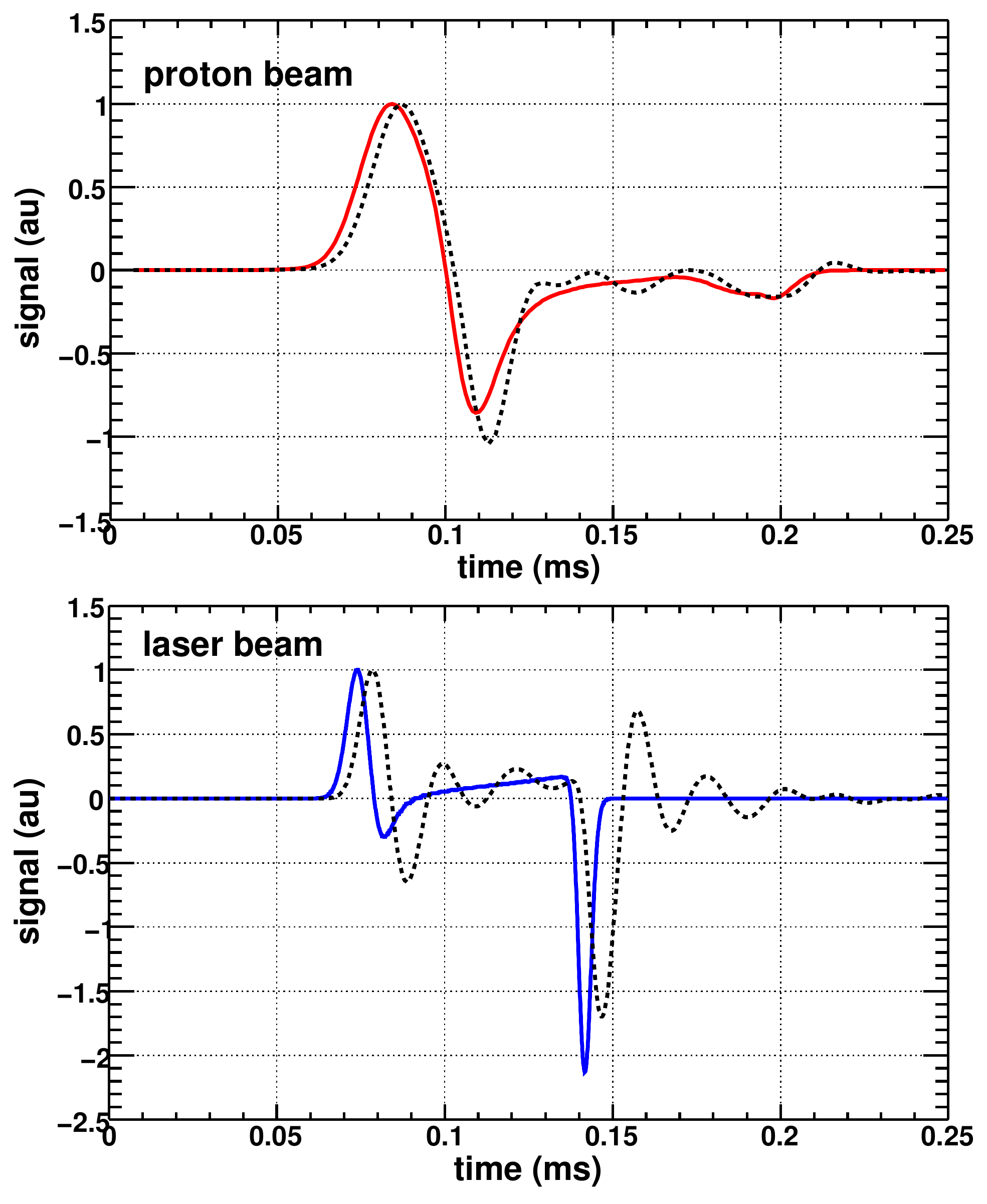}
    \caption{Simulated signals for the proton (upper graph) and laser experiment (lower graph). Both signals (solid lines) were calculated for the experimental conditions of Fig.\ \ref{fig_signals}. The corresponding dashed-lined signals mark the simulated signals after convolution with the sensor response. For better comparability, the signal maxima were normalised to 1.
      \label{fig_signals_sim} }
  \end{center}
\end{figure}

A series of simulations was conducted, where the input parameters were
varied individually in the range given by the experiment, including
uncertainties. Especially the spatial and temporal beam profile have a
substantial impact on amplitude, duration and shape of the signal. Simulated
signals and the respective sensor response corresponding to the
measured signals of Fig.\ \ref{fig_signals} are shown in Fig.\
\ref{fig_signals_sim}.  To minimise systematic effects from the setup
caused e.g.\ by reflections on the surfaces, the sensor response was
convoluted onto the simulated signals, rather than deconvoluted from
the measured ones. Thus in the analysis voltage rather than pressure signals are compared.

The shapes of the simulated pulses are altered by the sensor
response, especially the high-frequency components above the resonant
frequency of the sensor.  In the case of the laser pulse, mostly the
resonance of the sensor is excited, leading to a strong ringing. For
the proton beam, the primarily bipolar shape is again prominent,
whereas the laser pulse is segmented into the two parts described above.
The direct pulse of the laser experiment exhibits a bipolar shape as
well, albeit less symmetric than for the proton beam.

\begin{figure}[htb]
  \begin{center}
    \subfigure[Proton beam simulation results]{\parbox[c]{8.6cm}
      {\includegraphics[width=8.6cm]{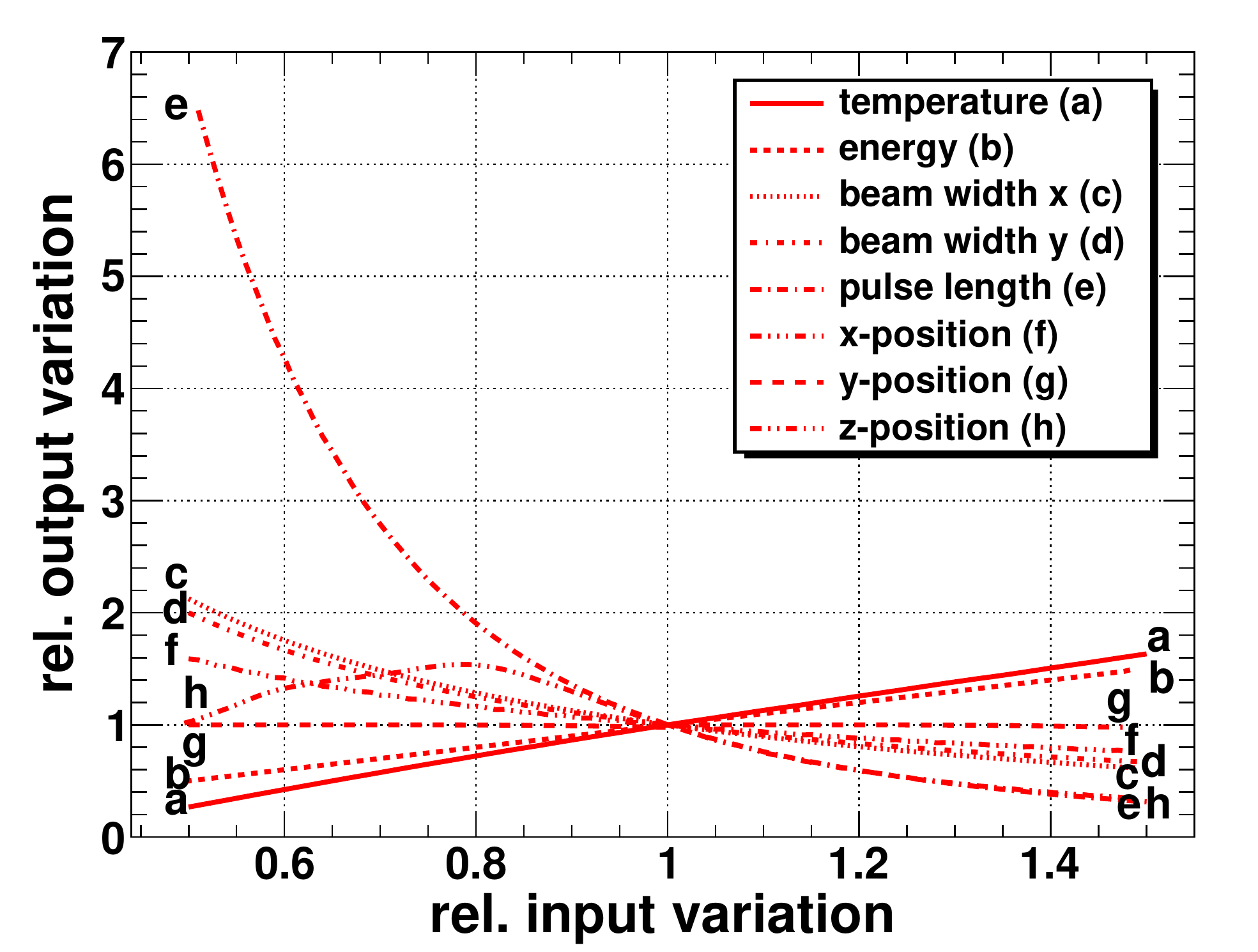}\label{fig_systematics_prot}}}
    \hspace{\subfigtopskip} \subfigure[Laser beam simulation
    results]{\parbox[c]{8.6cm}
      {\includegraphics[width=8.6cm]{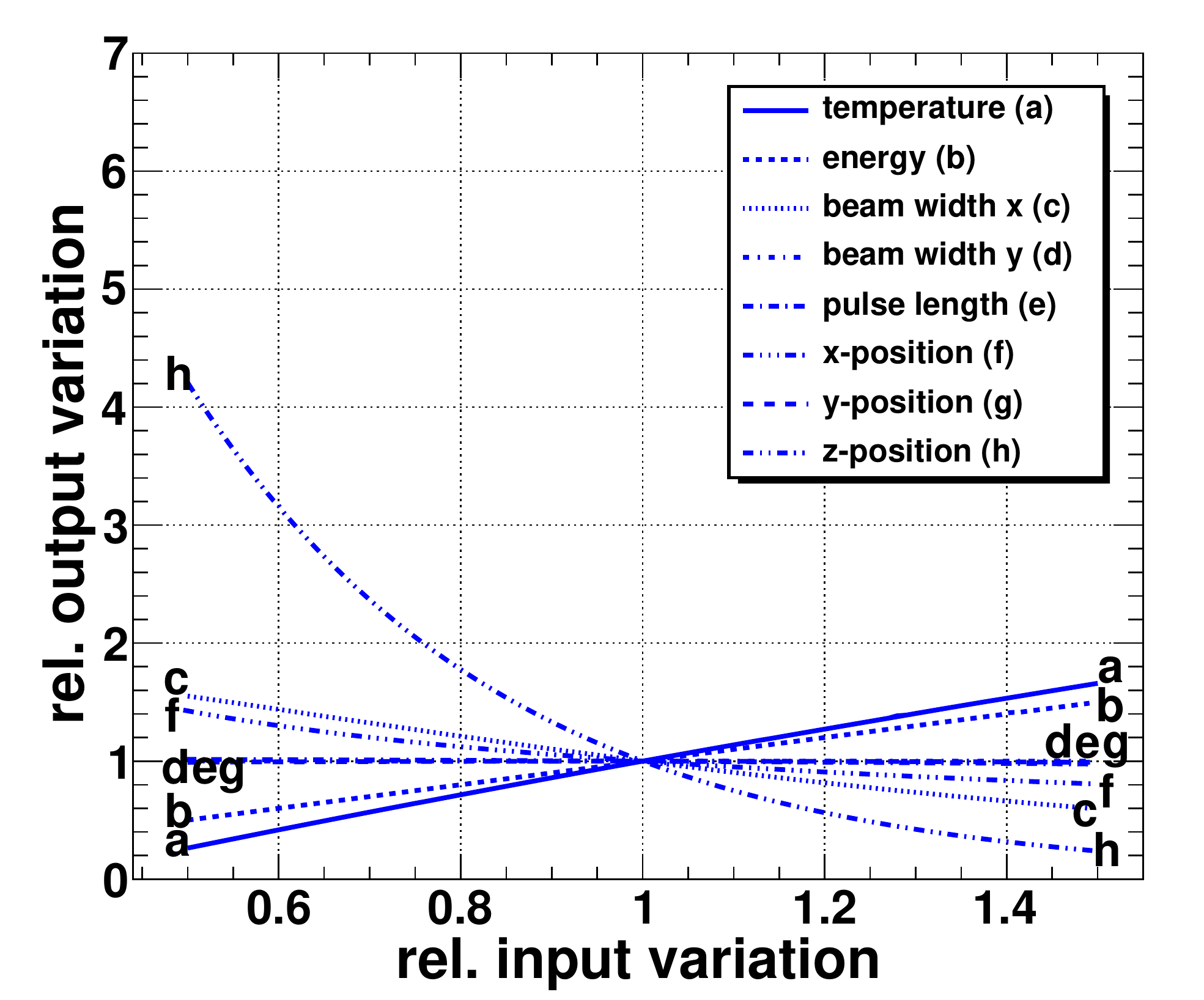}\label{fig_systematics_laser}}}
    \caption{Dependency of the simulated signal amplitude on variation of the simulation input parameters (labelled a - h) for the proton (upper graph) and the laser experiment (lower graph). With
    a nominal position $y=0$\,m of the sensor, the full range
    of the variation corresponds to $\pm2$\,cm for this parameter.
      \label{fig_systematics} }
  \end{center}
\end{figure}

Figure \ref{fig_systematics} exemplifies the dependency of the signal
amplitude on the input parameters of the simulation: water
temperature, pulse energy, beam profile in $x$ and $y$, pulse length and
the position of the sensor. As nominal positions of the sensor
$x=0.10\,$m, $y=0.0$\,m, and $z=0.22\,$m were used. 
All parameters were varied by $\pm 50\%$
around the value of the best agreement with measurement,
i.e.\ the values used for the simulations of
the signals shown in Fig.\ \ref{fig_signals_sim}.

Some of the characteristics of thermo-acoustic sound generation are
observable.  
As discussed in Sec.~\ref{sec_model},
the dependence on temperature enters through the factor $v_s^2\, \alpha/c_p$
(where the dependence on the speed of sound is negligible) 
and is roughly linear in the range investigated for this study.
The
dependence on energy is strictly linear. 
The dependency on the beam pulse
parameters is diverse. It is governed by the integral in Eq.\
(\ref{eq_pressure}) and therefore depends on both the spatial and the temporal
beam profiles and the interaction of the particles with the water. For
a given point in space and time,
the elementary waves produced
in the volume of energy deposition
may interfere constructively or
destructively depending on the beam properties. 
Accordingly, the length of the laser pulse has
no influence on the amplitude, as with 9\,ns it
is much shorter than the transit time of the acoustic signal through the
energy deposition area. For the several ten $\upmu$s long proton
spill, the spill time is comparable to the transit time. Thus the
acoustic signal shows a strong dependence on the spill time.  
The dependence on the radial coordinate $r$ w.r.t. the beam axis $r$ 
follows roughly the expected 1/$\sqrt{r}$ and
$1/r$ fall-off of a cylindric source in the near and the far field,
respectively. The $y$-position was varied between $-$2 and 2\,cm, as the
signals were recorded within the $xz$-plane. The resulting change in
amplitude is below 1\%. The $z$-dependency for the laser experiment
follows the exponential fall-off expected from the light absorption.
The one for the proton experiment shows the only non-strictly monotonic
behaviour due to the form of the energy deposition with the prominent
Bragg peak.

Using these dependencies the systematic uncertainties of the model 
were obtained using the experimental uncertainties of the various
parameters.  The main uncertainty for the proton beam is given by the
temporal profile of the pulse, which was simplified to a
Gaussian profile for this chapter (however not for the rest of this
work). For the laser beam experiment this parameter influences the
signal amplitude only on a one percent level. The second main
influence on the amplitude is the sensor position along the beam axis 
($z$-direction).

Table \ref{tab_systematics} gives the parameters and their
uncertainties ($\Delta V$) used for the simulation of the signals. The
resulting systematic uncertainties in the amplitude ($\Delta A$) are
given as well. The combined uncertainties are $^{+34\%}_{-29\%}$ for
the proton signal and $^{+26\%}_{-24\%}$ for the laser signal,
respectively.

\begin{table}[t]
\centering
  \begin{tabular}[h]{lcccc}
    Parameter      & \multicolumn{2}{c}{Proton}                                   &  \multicolumn{2}{c}{Laser}      \\
                   & value ($\Delta V$)       & $\Delta A$               &  value ($\Delta V$)         & $\Delta A$ \\[1mm] \hline \hline
  \mr{temperature} & $12.0^{\circ}$C          & \mr{$^{+10\%}_{-11\%}$}  & $12.0^{\circ}$C             & \mr{$\pm 11\%$}\\[-0.5mm]
                   & ($\pm 8\%$)              &                          &  ($\pm 8\%$)                & \\ [1mm]
  \mr{energy}      & $410$\,PeV   & \mr{$\pm 10\%$}          & $520$\,PeV      & \mr{$\pm 10\%$}\\[-0.5mm]
                   & ($\pm 10\%$)             &                          &  ($\pm 10\%$)               & \\[1mm] 
    beam width     & $3.0$\,mm                & \mr{$^{-14\%}_{+19\%}$}  & $5.0$\,mm                   & \mr{$^{-14\%}_{+16\%}$}\\[-0.5mm]
    ($x$-direction)  & ($\pm 14\%$)             &                          &  ($\pm 15\%$)               & \\ [1mm]
    beam width     & $3.0$\,mm                & \mr{$^{-12\%}_{+16\%}$}  & $5.0$\,mm                   & \mr{$\pm 0\%$}\\[-0.5mm]
    ($y$-direction)  & ($\pm 14\%$)             &                          &  ($\pm 15\%$)               & \\ [1mm]
  \mr{pulse length}& $1.0 \cdot 10^{-5}$\,s   & \mr{$^{-13\%}_{+16\%}$}  & $6.9 \cdot 10^{-7}$\,s      & \mr{$^{-2\%}_{+2\%}$}\\[-0.5mm]
                   & ($\pm 11\%$)             &                          &  ($\pm 50\%$)               & \\ [1mm]
  \mr{$x$-position}  & $0.10$\,m                & \mr{$^{-4\%}_{+5\%}$}    & $0.11$\,m                   & \mr{$^{-3\%}_{+4\%}$}\\[-0.5mm]
                   & ($\pm 7\%$)              &                          &  ($\pm 6\%$)                & \\ 
  \mr{$y$-position}  & 0.0\,m                   & \mr{$^{-2\%}_{+0\%}$}    & 0.0\,m                      & \mr{$^{-1\%}_{-1\%}$}\\[-0.5mm]
                   & ($\pm 1$\,cm)            &                          &  ($\pm 1$\,cm)              & \\ [1mm]
  \mr{$z$-position}  & 0.22\,m                  & \mr{$\pm 9\%$}           & $0.17$\,m                   & \mr{$^{-11\%}_{+13\%}$}\\[-0.5mm]
                   & ($\pm 4\%$)              &                          & ($\pm 5\%$)                 & \\ 
\hline
\mr{combined}           &   & \mr{$^{+34\%}_{-29\%}$} & &  \mr{$^{+26\%}_{-24\%}$}  \\
& & & & \\
  \end{tabular}
  \caption{Beam parameters used for the simulated signals in
    Fig.~\ref{fig_signals_sim} with their associated experimental
    uncertainties ($\Delta V$) and resulting uncertainty in signal
    amplitude ($\Delta A$). The pulse length of the laser is set to a
    value much higher than in the experiment, to save calculation
    time. The resulting uncertainty in the signal amplitude is below
    $2\%$}
  \label{tab_systematics}
\end{table}

\section{Analysis Results}
In the following, the results obtained in the comparison of
experiment and simulation are presented.

\subsection{Comparison of Simulated and Measured Signals}

\subsubsection{Proton Beam Experiment}\label{sec_signal_proton}
\begin{figure}
  \begin{center}
    \subfigure[Signals at $z=0.11$\,m]{\parbox[c]{8.6cm}
      {\includegraphics[width=8.6cm]{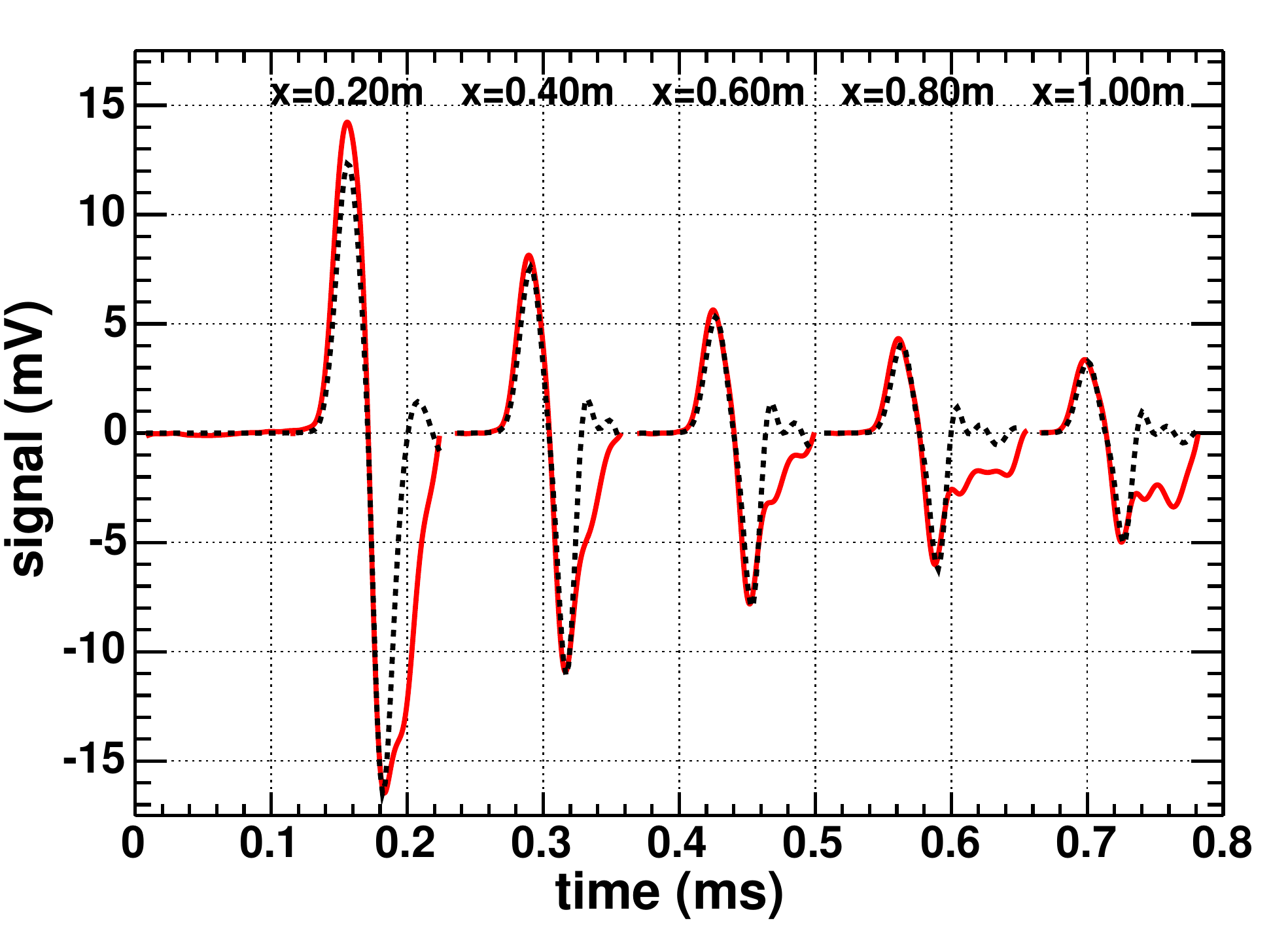}\label{fig_signal_sim1}}}
    \hspace{\subfigtopskip} 
    \subfigure[Signals at $z=0.21$\,m]{\parbox[c]{8.6cm}
      {\includegraphics[width=8.6cm]{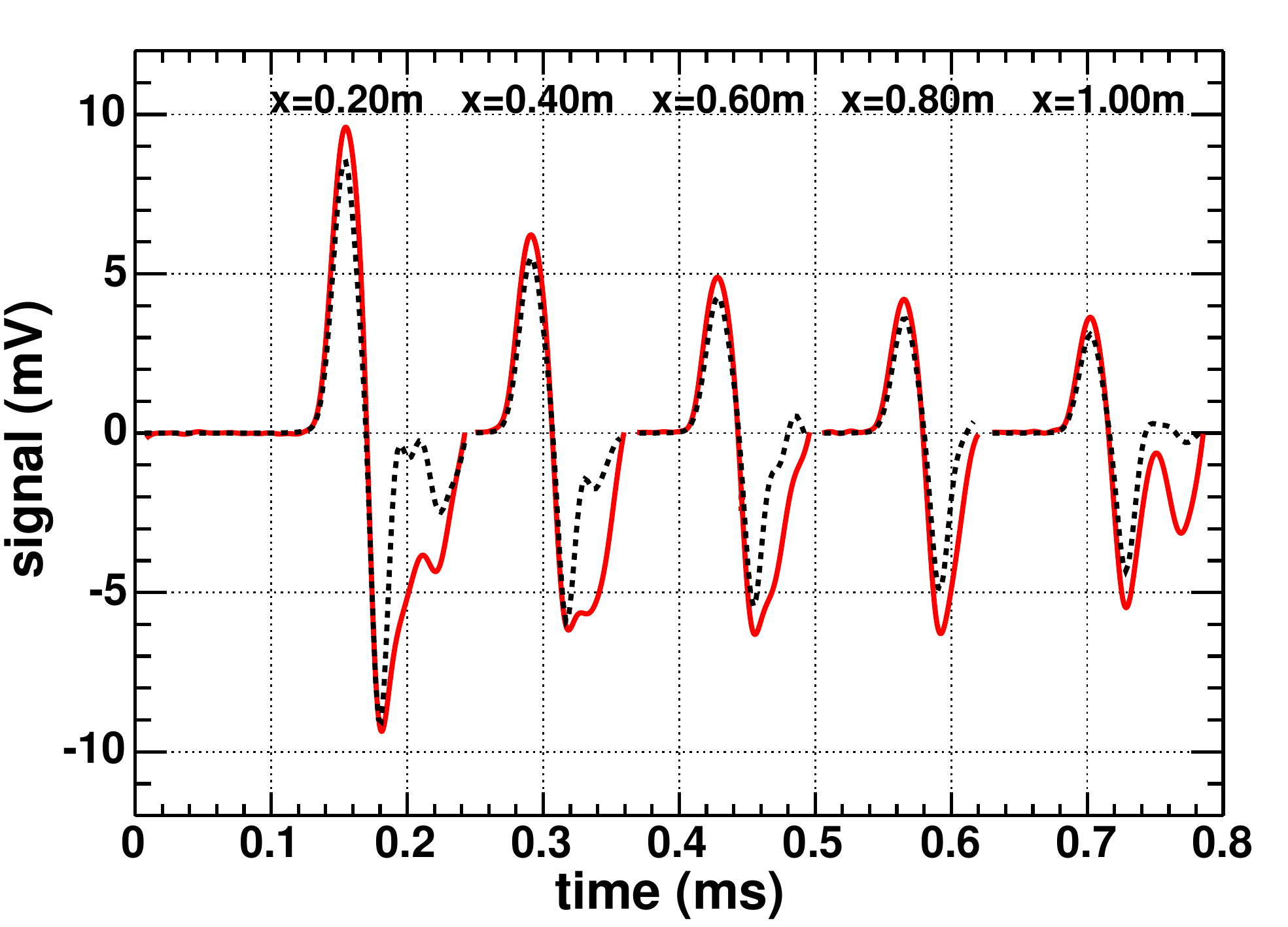}\label{fig_signal_sim3}}} 
    \hspace{\subfigtopskip}
    \subfigure[Signal at $x=0.01$\,m, $z=0.41$\,m]{\parbox[c]{8.6cm}
      {\includegraphics[width=8.6cm]{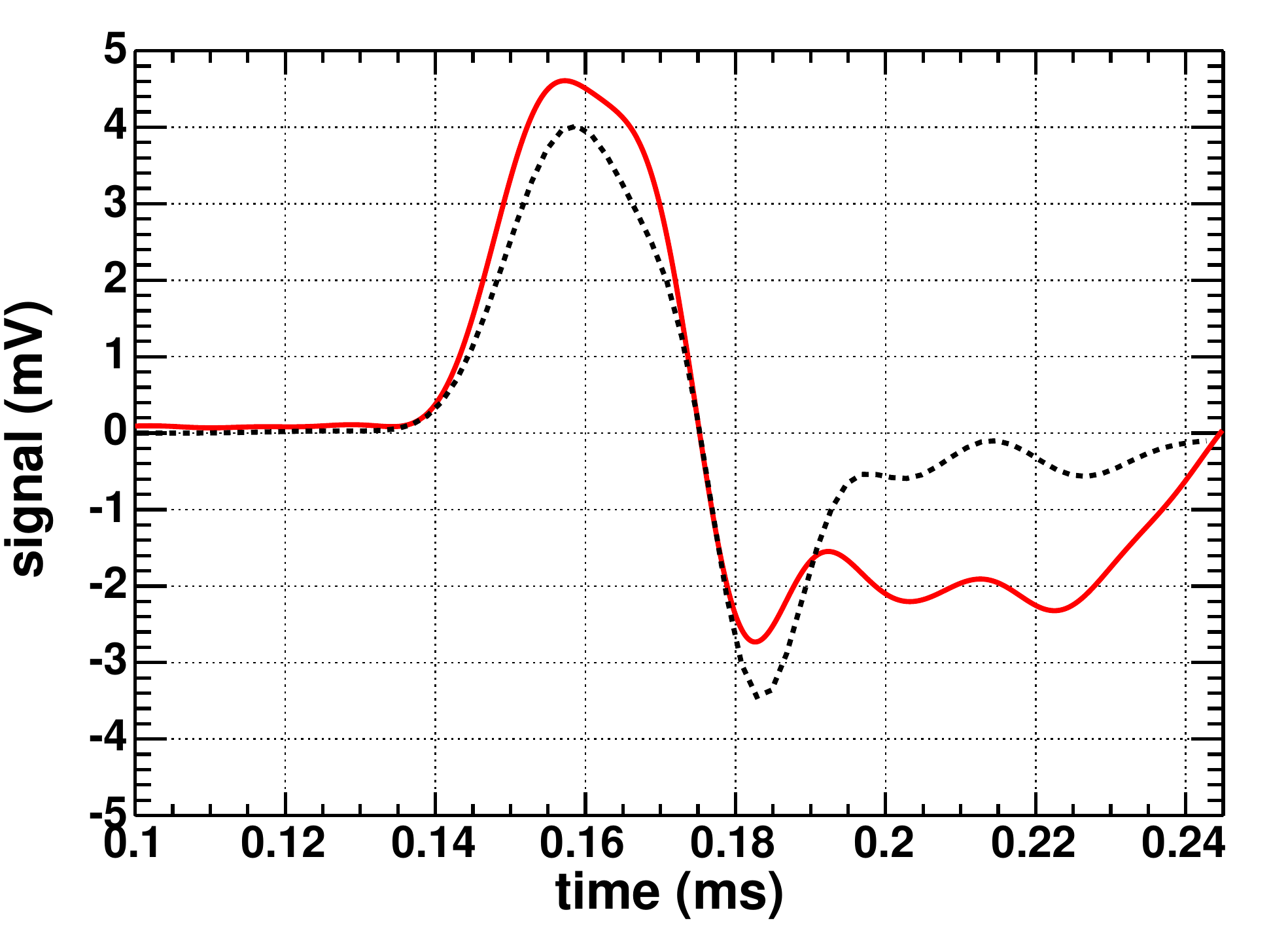}\label{fig_signal_sim4}}}
    \caption{Comparison of measured (solid line) and simulated (dashed line)
      signals at different sensor positions for the proton beam experiment. 
Note the different scale of the $x$-axis used in plot \subref{fig_signal_sim4}.     
      The reference signal for tuning the simulations is at $z=0.11$\,m and $x=0.40$\,m. The coordinate system is given in Fig.~\ref{fig_test_setup}.
      See text for a detailed description.
      \label{fig_proton_sim_meas} }
  \end{center}
\end{figure}

Figure \ref{fig_proton_sim_meas} shows a comparison between simulated
and measured signals for the proton beam experiment at different
sensor positions. 

For better visibility only the main part of the signals (first bipolar
part) is plotted.  The input parameters of the simulation were varied
within the experimental uncertainties until the best agreement with the measured signal in
amplitude and duration was obtained for the reference point at
$x$=0.40\,m and $z$=0.11\,m. 
For this optimisation, a simple procedure of adjusting the parameters manually and scanning the
resulting agreement  visually
was found to be sufficient.
The other signals were simulated with the same
parameter set, only the sensor positions were changed.  The signal
shapes differ for different $z$-positions due to the geometry of the
energy deposition profile described in Sec.\ \ref{sec_setup} with
cylindric form in the $xy$-plane and almost flat energy density in $z$-direction 
up to the Bragg peak at
$z$=0.22\,m. Due to this geometry an almost cylindrical wave is excited
in the medium, with coherent emission perpendicular to the beam axis.
In the region $z \lesssim 0.2$\,m the signals are of bipolar shape.
Along the beam axis ($x\approx0.0$\,m, $z > 0.25$\,m) the main part of
the observed signal originates from the Bragg peak as a nearly spheric
source and no clear bipolar shape evolves.

The agreement between simulation and measurement is good for all
positions. Not only amplitude and duration match (see also the
following sections) but also the signal shape is reproduced to a very
high degree.  The small discrepancies, primarily in the rare-faction
part of the bipolar pulse, have contributions stemming from a non-ideal sensor
calibration and reflections on the tank surfaces. However, a
significant part of the discrepancies may lie in the beam pulse
modelling or even the thermo-acoustic model itself.

\subsubsection{Laser Beam Experiment}\label{sec_signal_laser}
\begin{figure}[]
  \begin{center}
    \includegraphics[width=8.6cm]{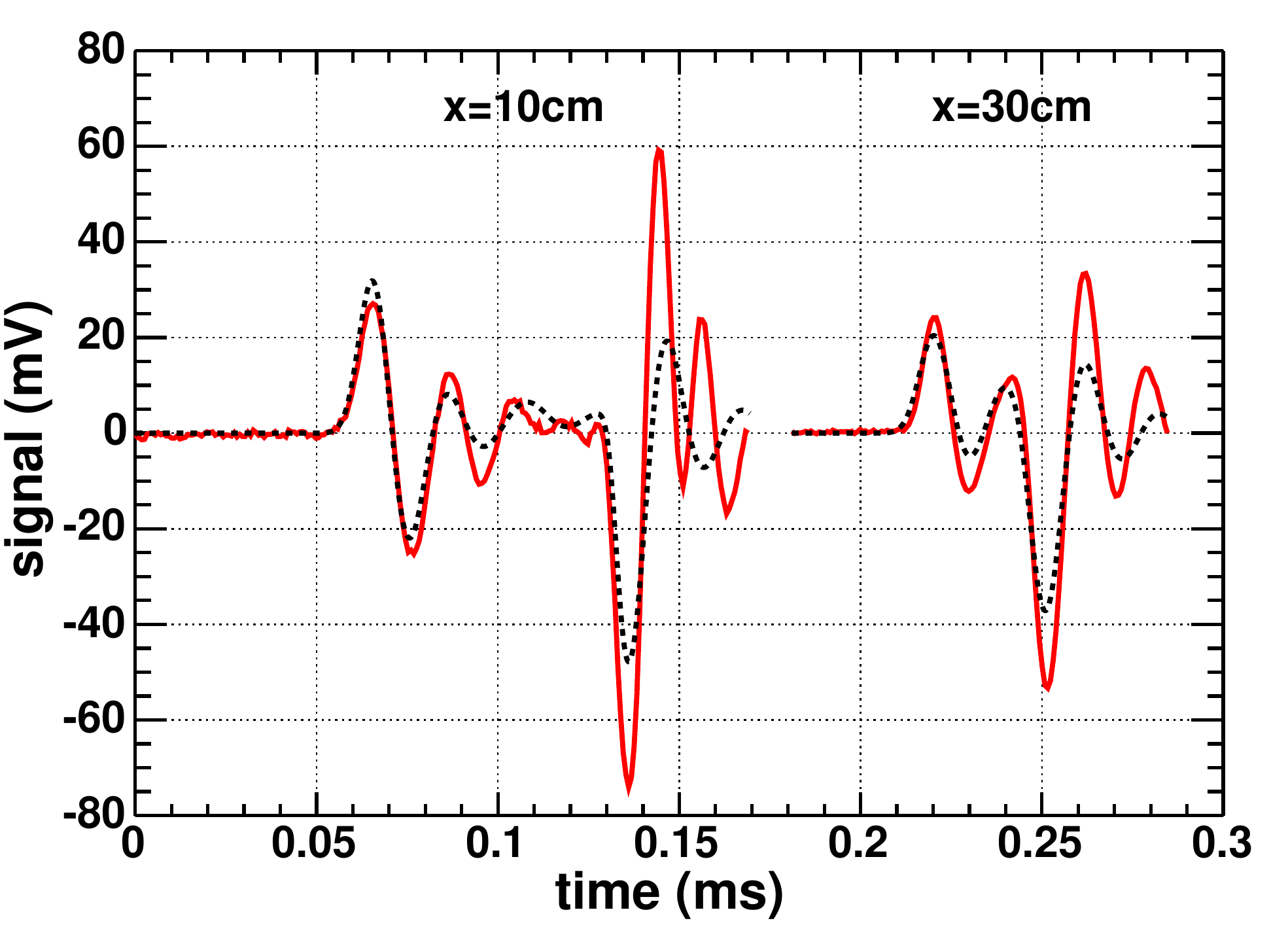}
    \caption{Comparison of measured (solid line) and simulated (dashed line) signals for the laser beam experiment at $z=0.17$\,m for different sensor positions along $x$ (reference at $x=0.10$\,m). For more details see text.
      \label{fig_laser_sim_meas} }
  \end{center}
\end{figure}

The prominent feature of the energy deposition of the laser beam is at
the beam entry into the medium. Overall, the geometry of this
deposition is mainly a cylindric one with rotational symmetry around the $z$-axis, 
as for the proton beam,
with coherent emission perpendicular to the beam axis (direct signal). The
signal from the discontinuity at the beam entry is emitted almost as
from a point source (beam entry signal). In contrast to the proton
beam, the shapes of signals at different positions along the $z$-axis
do not vary much, only the relative timing between the two signal
components varies. Therefore, Fig.\ \ref{fig_laser_sim_meas} shows only
signals for a $z$-position in the middle of the water tank. Here, both
signal parts are well described by the simulation. The beam entry part
of each signal is less well reproduced in the simulation due to its
high frequency components where the sensor calibration is less well
understood.
\\

To compare the characteristics of the signals in simulation and
measurement in the following studies, their amplitudes and the point in time of the recording of the 
signal maximum were further studied. For the laser
signal, the direct signal part is considered only.

\subsection{Speed of Sound Measurement}

\begin{figure}[htb]
  \begin{center}
    \includegraphics[width=8.6cm]{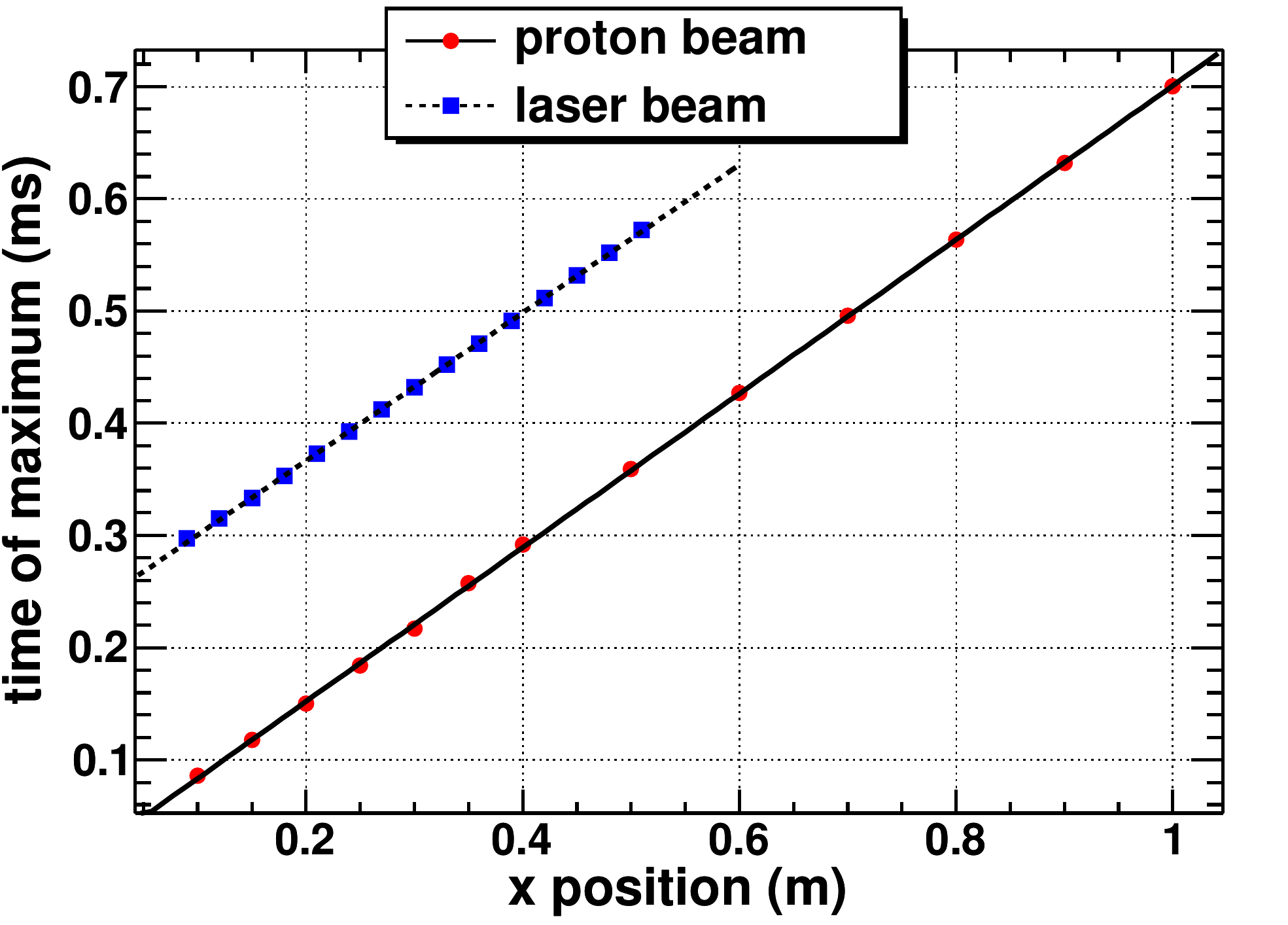}
    \caption{Arrival time of the signal maximum for different sensor
      position perpendicular to the beam ($x$ position). The straight lines
      represent linear fits to the data points yielding $v_s =
      1458\pm4\,{\textrm{m}}/{\textrm{s}}$ and $v_s =
      1508\pm3\,{\textrm{m}}/{\textrm{s}}$ for the proton and
      laser beam, respectively. \label{fig_time_vs_distance} }
  \end{center}
\end{figure}

Equation (\ref{eq_pressure}) yields as velocity of propagation for the
thermo-acoustic signal the speed of sound in the medium, here water.
To verify the hydrodynamic origin of the measured signals the
variation of the arrival times for different sensor positions
perpendicular to the beam axis were analysed. Figure
\ref{fig_time_vs_distance} shows the measured data and a linear fit
for each beam type. The data is compatible with an acoustic sound
propagation in water, as the fits yield a speed of sound compatible
with pure water at the temperature used. For the proton beam $v_s =
1458\pm4\,{\textrm{m}}/{\textrm{s}}$ ($\chi^2/\text{n.d.f.} = 8.2/11$) was obtained for a water temperature 
of $11.0\pm 0.2\,^\circ$C, where
literature \cite{DelGrosso} gives for pure water at normal pressure
$v_s = 1455\pm1\,{\textrm{m}}/{\textrm{s}}$
which is in complete 
agreement\footnote{The uncertainties of the theoretical values of the speed of sound result from the 
changing water temperature while the measurements were taken.}. 
For the
laser measurements the water temperature of $19.4\pm1.0\,^\circ$C and thus the speed of sound
were significantly higher, the observed $v_s =
1508\pm3\,{\textrm{m}}/{\textrm{s}}$   ($\chi^2/\text{n.d.f.} = 9.2/13$) is again in perfect agreement
with the theoretical value of
$1503\pm6\,{\textrm{m}}/{\textrm{s}}$. The offset between proton
and laser beam data in Fig.~\ref{fig_time_vs_distance} is due to a differing delay time in between trigger
time and arrival time of the different beams in the water
and is irrelevant  for the calculation of the speed of sound. 

\subsection{Variation of the Sensor Position}
The durations of the complete, unclipped signals
vary with sensor positions mainly due to reflections, which
were not simulated and can therefore not be compared.

\begin{figure}[htb]
  \begin{center}
    \includegraphics[width=8.6cm]{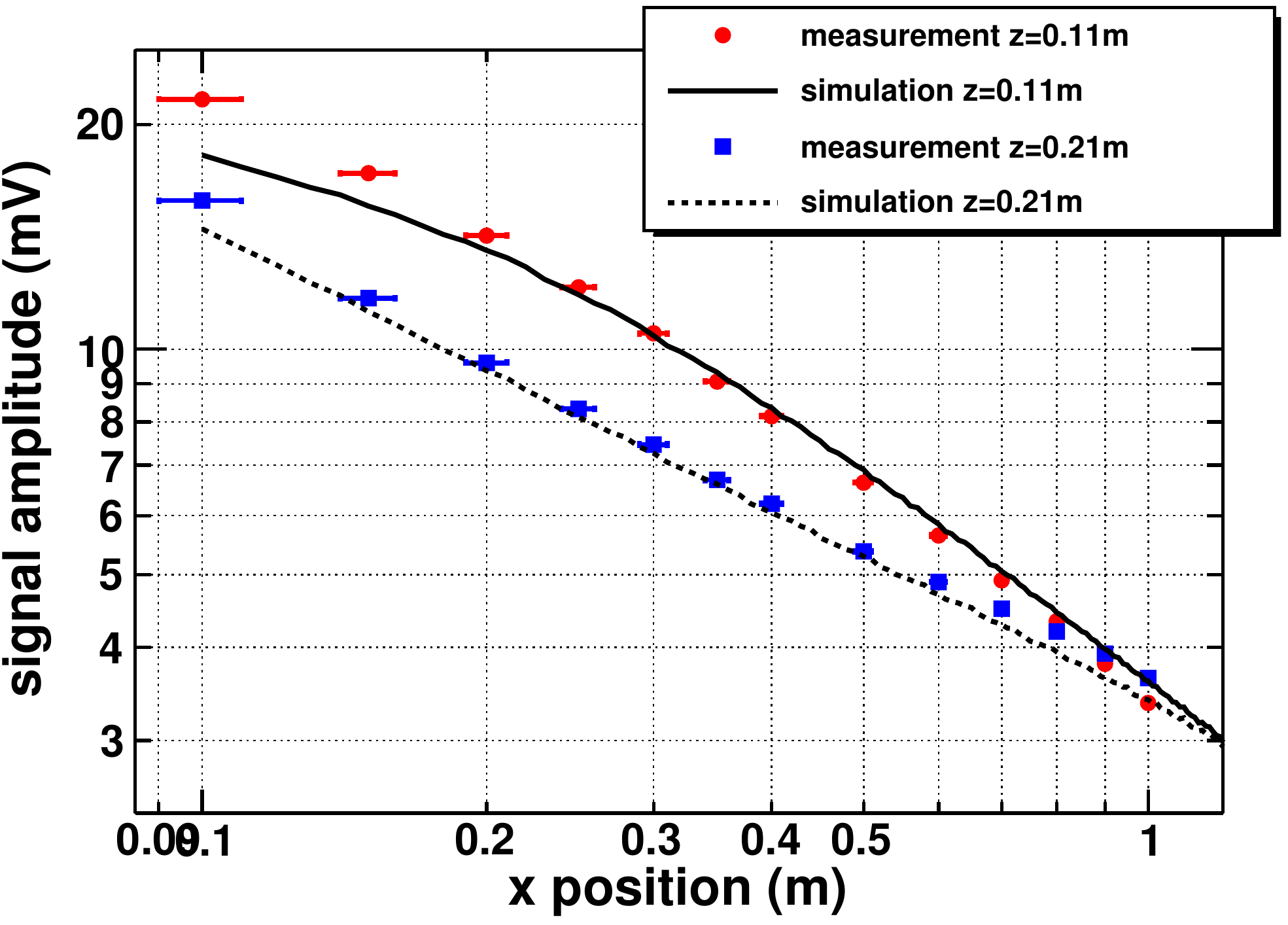}
    \caption{Development of the leading peak (maximum) with sensor
      position perpendicular to the beam axis ($x$) for the proton beam experiment. Points mark the
      measured amplitudes for two different $z$-positions and the lines
      the respective simulation results. \label{fig_amp_vs_x_protons}
    }
  \end{center}
\end{figure}

The comparison of simulated and measured signals for different sensor
positions within the water tank, excluding the parts of the signals dominated by reflections, was shown in
Fig.~\ref{fig_proton_sim_meas}. The good agreement between model and
measurement also manifests itself in the development of the signal amplitude
with distance of sensors from the beam axis, shown in
Fig.~\ref{fig_amp_vs_x_protons}. To minimise systematic effects from
reflections, only the amplitude of the leading maximum is analysed.

Though there are sizeable deviations, the overall shape of the curve
is reproduced.  The behaviour is again different for the two
$z$-positions. The development at $z=0.11$\,m follows the one expected
from a cylindric source with a $1/\sqrt{r}$ behaviour in the near
field up to $\sim$\,$0.3$\,m and a $1/r$ behaviour in the far field
beyond that distance. At the smallest measured distances, the
simulated behaviour deviates from the measured one. 
Presumably this is due to simplifications
made in the derivation of the model in Sec.~\ref{sec_model}. 
At $z=0.21$\,m the amplitude falls off more uniformly, this is
again a combination of the point-like emission characteristic of the
Bragg peak interfering with the cylindric emission at $z<0.2$\,m.

\begin{figure}[ht]
  \begin{center}
    \includegraphics[width=8.6cm]{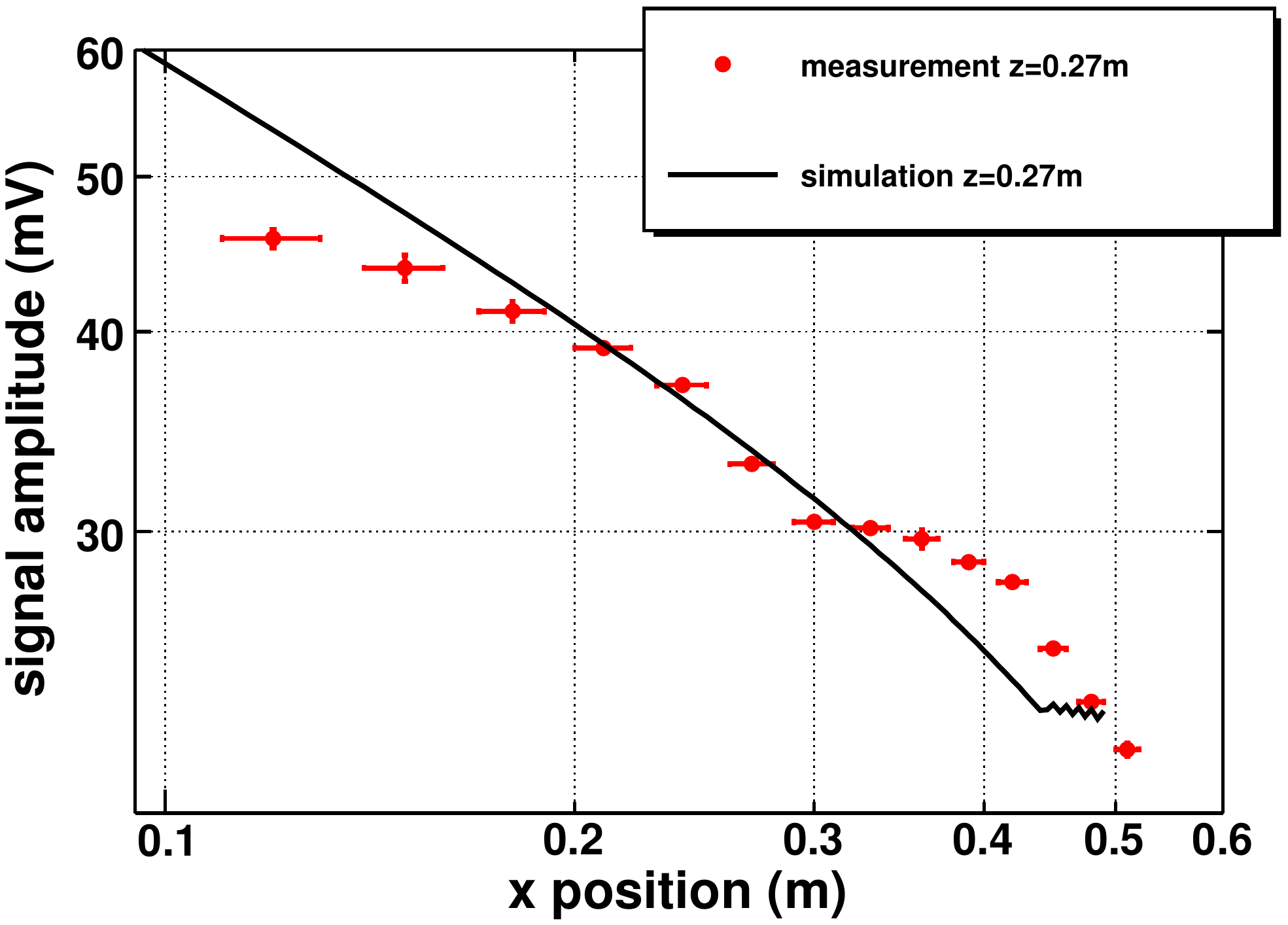}
    \caption{Development of the leading peak (maximum) with sensor
      position perpendicular to the beam axis ($x$) for the laser beam experiment. Points mark the
      measured amplitudes at $z=0.27$\,m and the line the simulation
      result. \label{fig_amp_vs_x_laser} }
  \end{center}
\end{figure}

The behaviour for the laser experiment is not as well reproduced by
the simulation (see Fig.~\ref{fig_amp_vs_x_laser}). This is again
attributed to the high-frequent signal components of the laser, where minor
uncertainties in the simulation
may result in big changes of the signal amplitude.
Especially the overlap of the direct with the beam entry window signal
distort  the signal shape. At distances exceeding 0.5\,m the two signal
parts cannot be distinguished.  

\subsection{Variation of the Pulse Energy}

\begin{figure}[htb]
  \begin{center}
    \includegraphics[width=8.6cm]{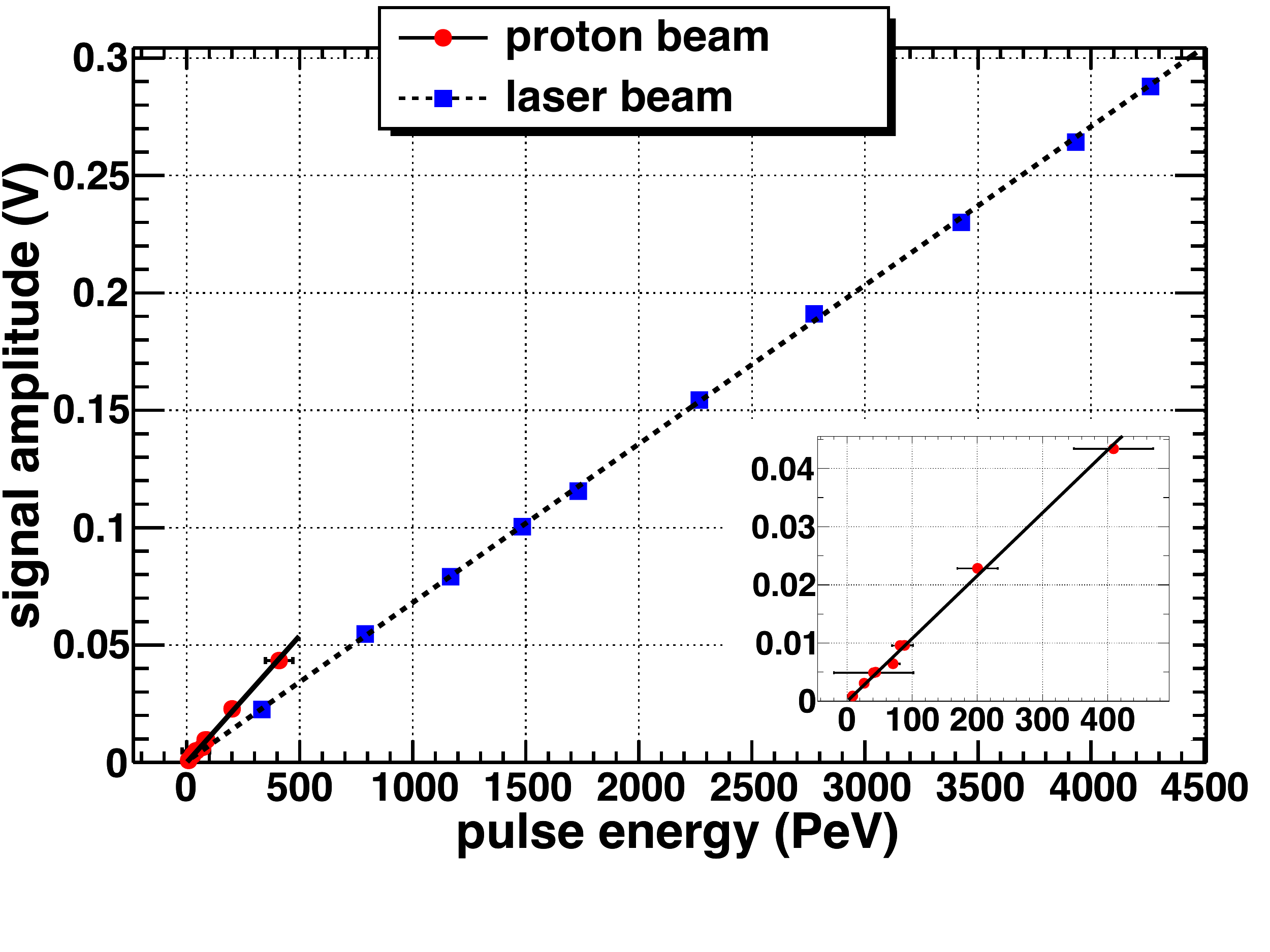}
    \vspace*{-5mm}
    \caption{Peak-to-peak amplitude of signals for different pulse
      energies, for both beams the sensors where at a position $x
      \approx 0.10$\,m. There is a ten percent systematic uncertainty
      in the absolute determination of the pulse energy.  The lines
      represent linear fits to the data points yielding a
      zero-crossing of the amplitude compatible with no energy in a
      pulse. The insert shows the data for the proton beam.\label{fig_amplitude_vs_energy} }
  \end{center}
\end{figure}

Assuming otherwise unchanged settings, the energy deposition density
$\epsilon$ scales linearly with total deposited energy. Thus the spill
energy can be written as a pre-factor in Eq.\ (\ref{eq_pressure})
effecting the pressure and thus signal amplitude linearly. As shown in
Fig.~\ref{fig_amplitude_vs_energy} this behaviour was observed in the
experiments yielding a zero-crossing of the pulse energy at
$(-4.1\pm5.3)\,$mPa for the proton beam and $(42\pm87)\,$mPa for the
laser beam. Both values are consistent with zero. The slope of the
line depends on the energy deposition and the sensor positioning along
the beam axis and can therefore not be compared for the two beams. 
As expected from the model, the
signal duration and signal shape showed no significant dependence on
energy. 

\subsection{Variation of the Temperature}
\label{subsec:tempvar}

\begin{figure}[hbt]
  \begin{center}
    \includegraphics[width=8.6cm]{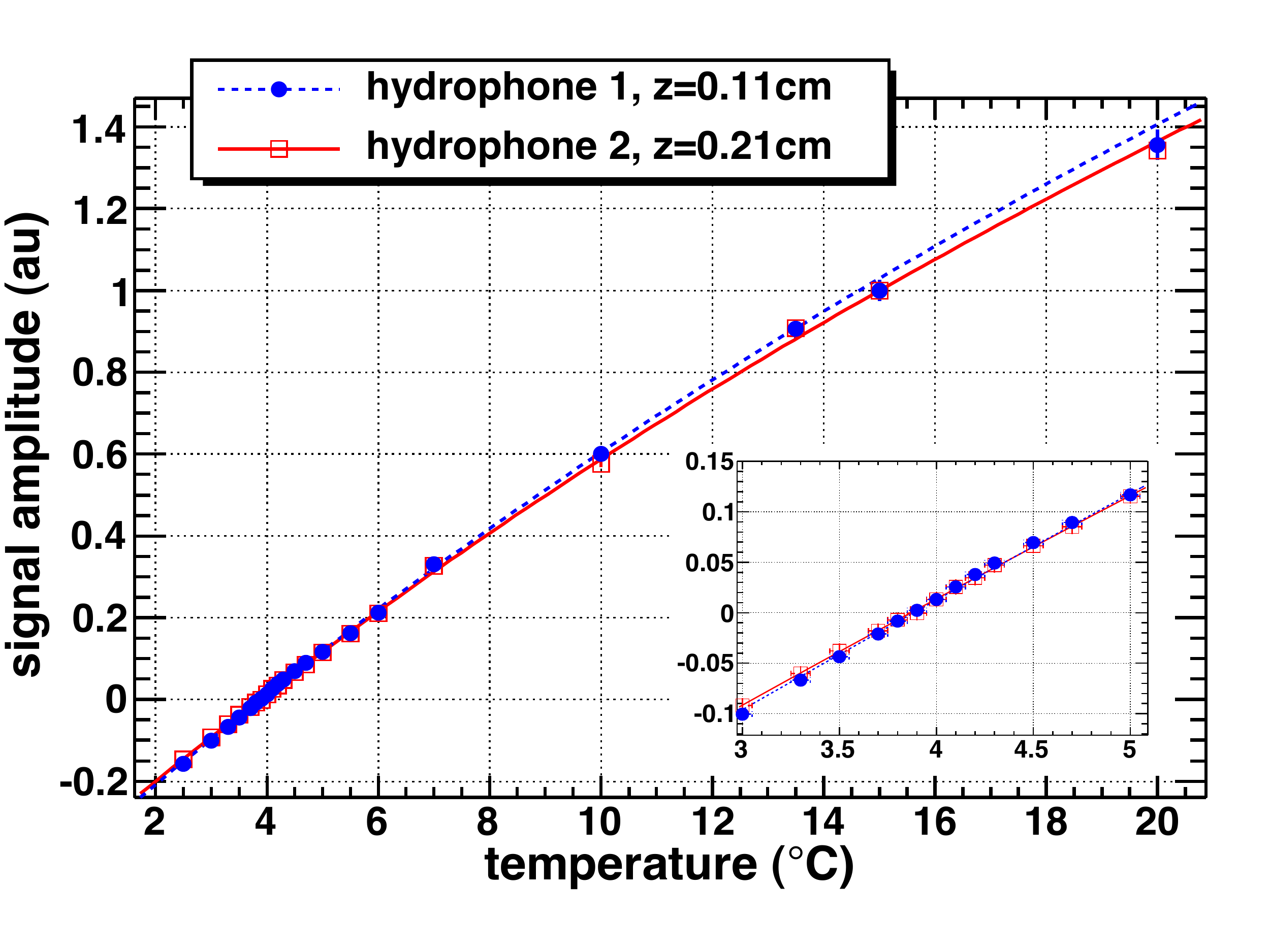}
     \vspace*{-4mm}
    \caption{ Measured signal amplitude of the bipolar acoustic signal
      induced by laser pulses at different temperatures fitted with
      the theoretical expectation as described in the text. All amplitudes
      were normalised to $1$ at
      $15.0^{\circ}\mathrm{C}$.
	  The insert shows a blow-up of the region around $4^\circ$\,C where the sign of the amplitude changes.      
      \label{fig_temp_laser}}
    \vspace{5pt} 
\end{center}
\end{figure}

The main feature of the thermo-acoustic model is its dependence on the
temperature of the medium.  Figure~\ref{fig_temp_laser}  shows the 
temperature dependence of the signal
peak-to-peak amplitude for the laser beam, where a positive (negative) sign denotes a
leading positive (negative) peak of the signal. The two data sets
shown in the figure were recorded by two sensors simultaneously,
which were positioned at $x=0.10\,\mathrm{m}$ perpendicular to the
beam axis and at $z=0.11\,\mathrm{m}$ and $z=0.21\,\mathrm{m}$ along
the beam axis, respectively. In the case of the proton beam setup, which will be discussed
below, these
hydrophone positions correspond roughly to the $z$-position of the
Bragg-peak and a $z$-position half way between the Bragg-peak and the
beam entry into the water, respectively. For comparability, the same
positions and the same sensors were chosen for the laser and proton beam experiments.

The laser beam signal shown in Fig.~\ref{fig_temp_laser} changes its
polarity around $4 ^\circ \mathrm{C}$, as expected from the
thermo-acoustic model. The theoretical expectation for the signal
amplitude, which is proportional to $\alpha/c_p$ and vanishes at
$4^\circ\mathrm{C}$ for the given temperature and pressure, is fitted to the experimental data. In the fit,
an overall scaling factor and a shift in temperature (for the
experimental uncertainty in the temperature measurement) were left
free as fit parameters.  The fit yielded a zero-crossing of the
amplitude at $(3.9 \pm 0.1)^\circ \mathrm{C}$, where the error is
dominated by the systematic uncertainty in the temperature setting.

\begin{figure}[hbt]
\begin{center}
    \includegraphics[width=8.6cm]{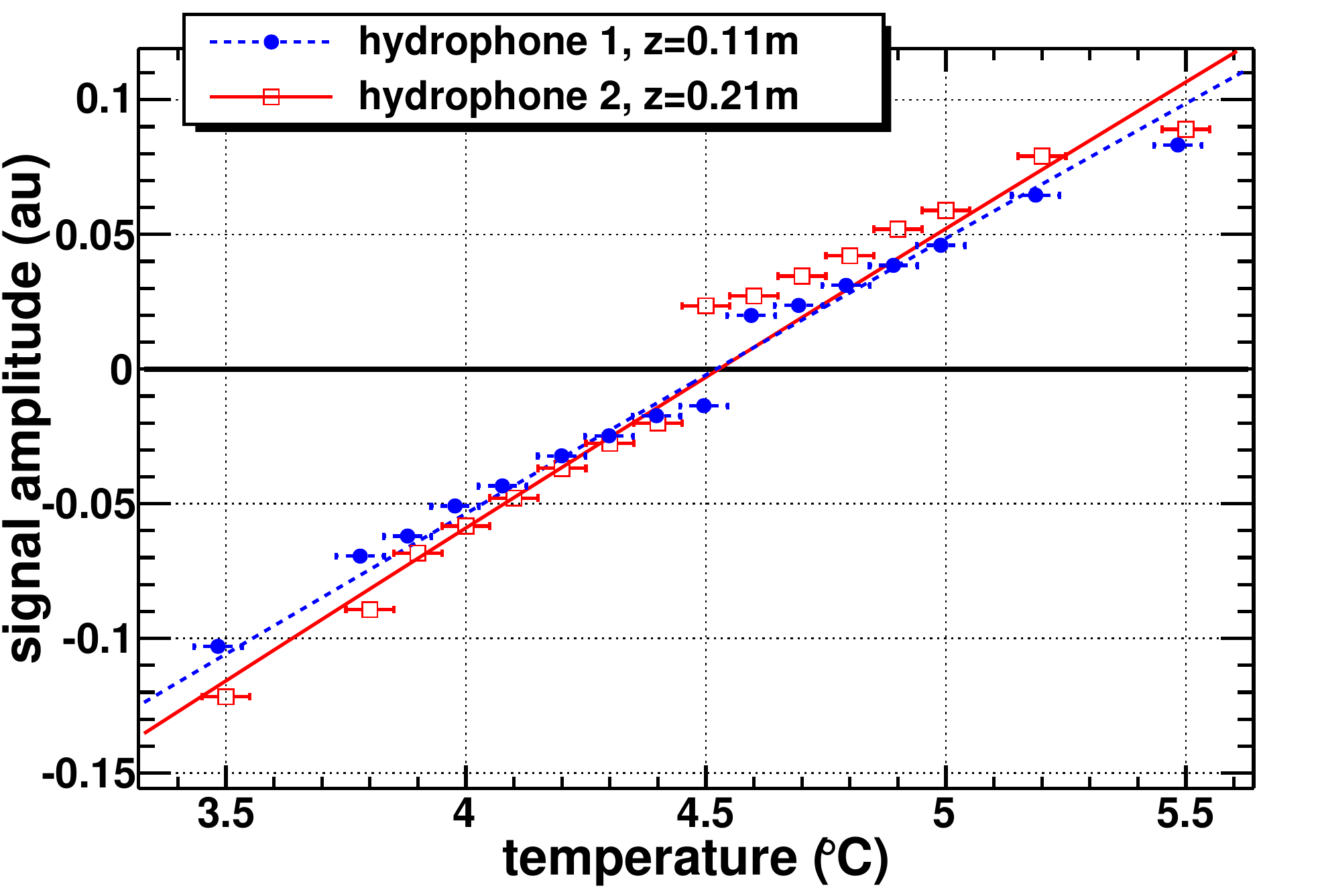}
    \caption{
    Measured signal amplitude of the bipolar acoustic signal
      induced by proton pulses at different temperatures near 4.0$^{\circ}$C, fitted with
      the model expectation as described in the text. A systematic deviation from the
      model expectation as the amplitude changes its sign is clearly visible. 
      The amplitudes were normalised to $1$ at $15.0^{\circ}\mathrm{C}$.           
      \label{temp_amp_protons_uncorr} }
\end{center}
\end{figure}

\begin{figure}[hbt]
\begin{center}
    \includegraphics[width=8.6cm]{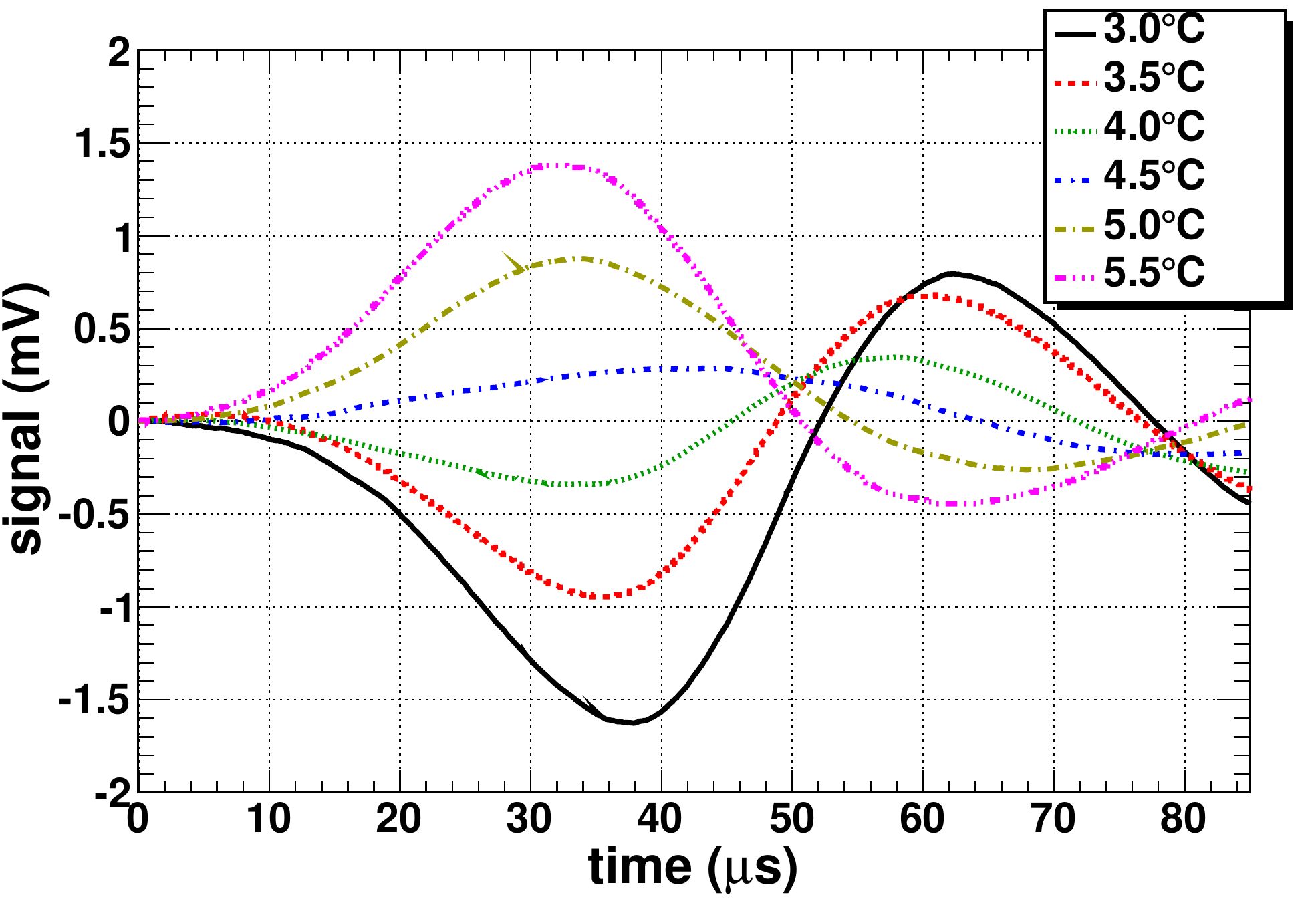}
    \caption{
Comparison of signals recorded for the proton beam with hydrophone 1 at different 
temperatures. For each temperature, the signals 
were first smoothed with a 100\,kHz
second order Butterworth low pass filter. Subsequently 1000 waveforms were averaged,
as described in Sec.~\protect\ref{sec_setup}. 
 To allow for an easy comparison of the signal shapes, a point in time at the onset of the acoustic signals was chosen as zero time and for all signal amplitudes the corresponding offset was added or subtracted to 
yield a zero amplitude at that time.         
      \label{temp_comp_signals} }
\end{center}
\end{figure}

\begin{figure}[bht]
\begin{center}
    \includegraphics[width=8.6cm]{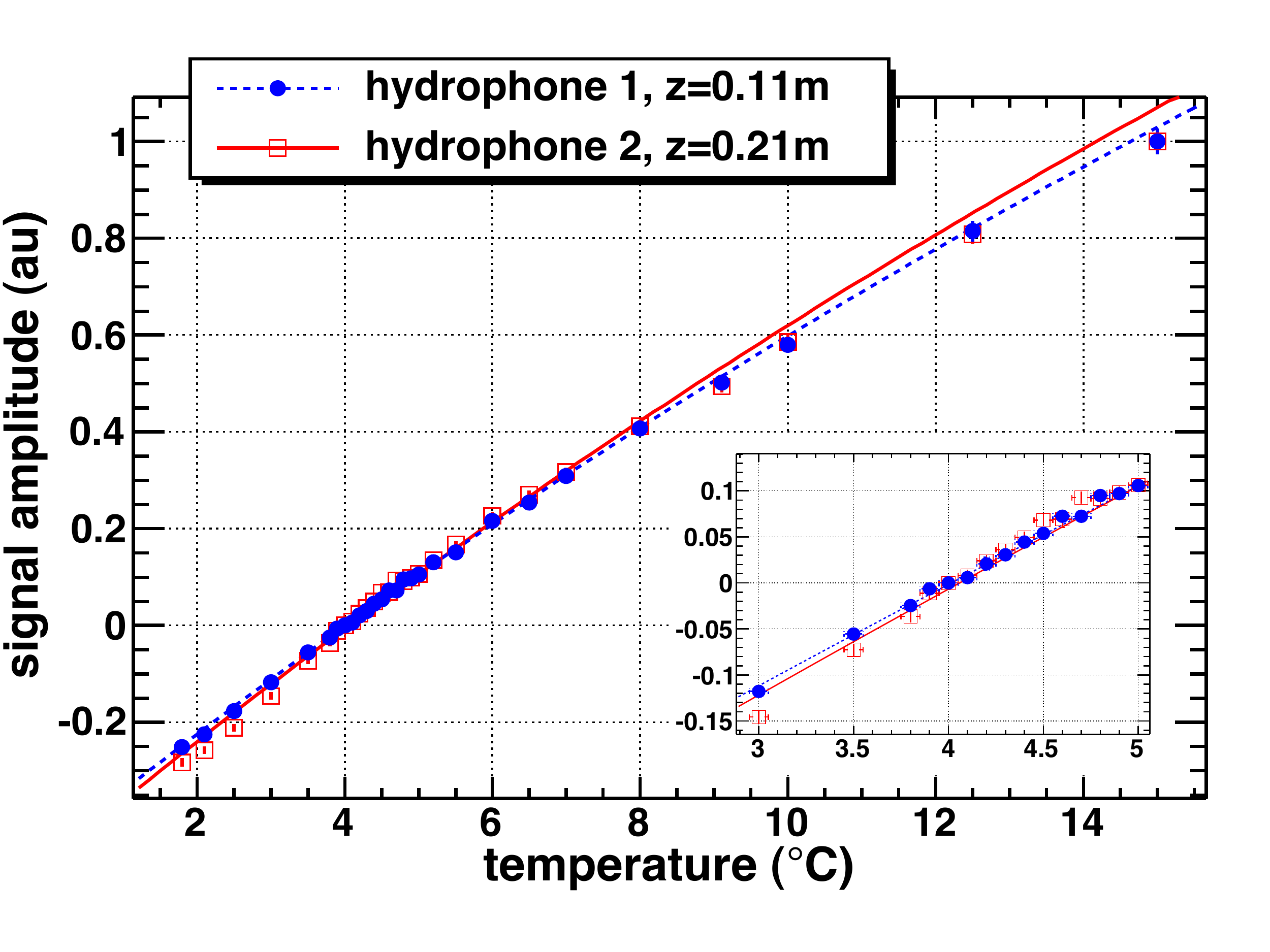}
     \vspace*{-4mm}
    \caption{ Measured signal amplitude of the bipolar acoustic signal
      induced by proton pulses at different temperatures fitted with
      the model expectation as described in the text. The
      non-thermo-acoustic signal at 4.0$^{\circ}$C was subtracted at
      every temperature. The amplitudes were afterwards normalised to
      $1$ at $15.0^{\circ}\mathrm{C}$.
	  The insert shows a blow-up of the region around $4^\circ$\,C where the sign of the amplitude changes.       
      \label{fig_temp_proton} }
\end{center}
\end{figure}

Analysing the proton data in the same fashion resulted in a fit that 
deviated from the model expectation, and a zero-crossing
significantly different from $4.0^\circ \mathrm{C}$ at $(4.5 \pm
0.1)^\circ \mathrm{C}$, see Fig.~\ref{temp_amp_protons_uncorr}. The data strongly indicate
 the presence of a systematic effect near the zero-crossing of the signal amplitude. 
To understand this effect, the signal shapes near the temperature of $4.0^\circ \mathrm{C}$ were investigated (Fig.~\ref{temp_comp_signals}). A non-vanishing signal is clearly observable at $4.0^\circ \mathrm{C}$ and
the signal inverts its polarity between $4.0^\circ \mathrm{C}$ and $4.5^\circ \mathrm{C}$.
In view of the results from the laser beam
measurements and the obvious systematic nature of the deviation from the model visible
in Fig.~\ref{temp_amp_protons_uncorr}, we subtracted the residual signal at
$4.0^\circ \mathrm{C}$, which has an amplitude of $5\%$ of the
$15.0^{\circ}\mathrm{C}$ signal, from all signals. Thus a
non-temperature dependent effect in addition to the thermo-acoustic signal
was assumed.  The resulting amplitudes shown in
Fig.~\ref{fig_temp_proton} are well described by the model prediction.

The production mechanism of the underlying signal at $4.0^\circ
\mathrm{C}$, which was only observed in the proton experiment, could
not be unambiguously determined with the performed measurements. From
the model point of view, the main simplification for the derivation of
Eq.\ (\ref{eq_pressure}) was to neglect all non-isotropic terms and
momentum transfer to the medium in the momentum density tensor
$\Pi_{ij}^B$ by setting $\beta=0$ in Eq.~(\ref{eq:ansatz}).  
As discussed in Sec.~\ref{sec_model}, 
dipole radiation could
contribute significantly near the disappearance of the volume
expansion coefficient for the case  $\beta \neq 0$.
Also other non-thermo-acoustic signal
production mechanism have been discussed in the literature which could
give rise to an almost temperature independent signal, see e.g.\
\cite{Askaryan2}.  The obvious difference to the laser experiment are
the charges involved both from the protons themselves and the
ionisation of the water which could lead to an interaction with the
polar water molecules. Another difference are the massive protons
compared to massless photons. Residual signals at $4.0^\circ
\mathrm{C}$ were found in previous works as well \cite{Sulak, Hunter1,
  Hunter2, Albul2},
as will be discussed in more detail in Sec.~\ref{subsec:comparison}.

For clarification further experiments are needed either with ionising
neutral particles (e.g.~synchrotron radiation) or with charged
particles (e.g.~protons, $\upalpha$-particles) with more sensors
positioned around the Bragg-peak. With such experiments it might be
possible to distinguish between the effect of ionisation in the water
and of net charge introduced by charged particles.

\subsection{Comparison with Previous Experiments}\label{subsec:comparison}

With the analysis that has been described above, the signal production according to the thermo-acoustic model could be unambiguously confirmed. While the model has been confirmed in previous experiments,
the simulations presented in this work constitute a new level of precision.
The most puzzling feature, a residual signal at 4$^\circ$C that was also observed in previous experiments, was investigated with high precision by scanning the relevant temperature region in steps of $0.1^\circ$C. 
The observed shift of the zero-crossing of the amplitude towards 
values higher than 4$^\circ$C, caused by a 
leading rarefaction non-thermal residual signal at  4$^\circ$C,
is in qualitative agreement with~\cite{Hunter1}.  
In \cite{Sulak},  a residual signal at $4^\circ$C was also reported.
Since in that work the zero-crossing of the amplitude is observed at $6^\circ$C, i.e.\
a higher value than the expected $4^\circ$C, it can be assumed that the corresponding residual signal has a leading rarefaction. 

In~\cite{Albul2}, a residual signal is found at $4.25^\circ$C, however with a leading compression rather than rarefaction. The authors conclude that this may lead to a signal disappearance point {\em below} the expected
value, in contrast to \cite{Hunter1, Sulak} and the work presented in this article. 

For the measurements with a laser beam reported in~\cite{Hunter2}, a residual signal was also observed at $4^\circ$C, albeit with a leading compression and a 
subsequent reduction of the temperature of the zero-crossing of the signal amplitude to about $2.5^\circ - 3.0^\circ$C. This observation is in contrast to the laser experiment presented in this article.

In conclusion, the works of all authors discussed here indicate a non-thermal residual signal for proton beams, albeit with varying results concerning the size of the effect and the shape of the underlying non-thermal signal.
The results for the laser beam reported in~\cite{Hunter2} differ from those described in the article at hands.
It should be pointed out, however, that in \cite{Hunter1} and \cite{Hunter2} results are reported by the same authors for proton and for laser beams, respectively. A comparison of these two publications shows that a different behaviour near
the temperature of $4^\circ$C was observed for the two types of beams.  
Hence, to the best of our knowledge, there are currently no results in contradiction with the 
notion of different non-thermal effects in the interaction of proton vs.\ laser beams with water.
The available data does not allow for a more detailed analysis of the correlation between experimental conditions 
and the temperature of the zero-crossing of the signal amplitude.

\section{Applications in Astroparticle Physics}
Efforts to detect neutrinos at ultra-high energies are at the frontier of research in the field of astroparticle physics. 
Neutrinos are the only viable messengers at ultra-high energies beyond the local universe, i.e.\ distances well beyond several tens of Megaparsecs.
If successful, the investigation of these elusive particles will 
not only enhance the understanding of their own nature, but also
provide important complementary information on the astrophysical phenomena and the environments that accelerate particles to such extreme energies. 

For acoustic particle detection, not only the technical aspects such as optimal
design and detector layout are subject of research. But also the underlying physics processes---i.e.\ the formation of hadronic cascades resulting from neutrino  interactions in dense media---have never been observed directly in detector
experiments at these energies. Producing ever more reliable extrapolation of reaction properties to ultra-high energies is an ongoing effort. 
With  advancements in the simulations of cascades forming in water and improvements of  detector
simulation tools, the discrepancy between cascade parameters from independent simulations decreased: recent studies differ only slightly \cite{Bevan, aco2, niess, saund}. 
At the same time it is necessary to gain a solid understanding of the sound signals generated from the 
energy depositions by particle cascades. 
For this purpose, laboratory measurements are required.

This work, together with others \cite{Sulak, Hunter1, Hunter2, Albul1, Albul2, Capone, Demidov} has
established the validity of the thermo-acoustic model with
uncertainties at the $10\%$ level.  The input to a model as discussed in this article is an
energy deposition in water as it is also produced by a cascade that evolves from a neutrino
interaction. 
In comparison with the uncertainty in the thermo-acoustic model, 
 uncertainties due to the simulations of hadronic cascades
 and  cascade-to-cascade variations are large \cite{aco2},
dominating the challenge to detect and identify sound signals resulting from neutrino interactions.
It can hence be concluded that the current level of precision in modelling sound signals in the 
context of the thermo-acoustic model is fully sufficient for the understanding of acoustic neutrino signatures.
The latter is necessary to improve the selection efficiency and background rejection for neutrino detection algorithms 
in potential future acoustic neutrino detectors. 
Several experiments have been conducted~\cite{saund,Spats,Onde,Amadeus} to understand the acoustic background at the sites of potential future
large-scale acoustic neutrino telescopes in sea water, fresh water
and ice. 
The combination of 
simulation efforts, laboratory measurements and studies with in-situ test arrays will
allow for a conclusion of the feasibility of acoustic neutrino detection.

\section{Conclusions}
We have demonstrated that the sound generation mechanism of intense
pulsed beams is well described by the thermo-acoustic model.  In
almost all aspects investigated, the signal properties are consistent
with the model. The biggest uncertainties of the experiments are on the 10$\%$
level. One discrepancy is the non-vanishing signal at 4$^{\circ}$C for
the proton beam experiment, which can be described with an additional
non-temperature dependent signal with a $5\%$ contribution to the amplitude at
15$^{\circ}$C. The model allows for calculations of the
characteristics of sound pulses generated in the interaction of high
energy particles in water with the input of the energy deposition of
the resulting cascade. A possible application of this technique would
be the detection of neutrinos with energies $\gtrsim 1\,\mathrm{EeV}$.

\section*{Acknowledgements}
This work was supported by the German ministry for education and
research (BMBF) by grants 05CN2WE1/2, 05CN5WE1/7 and 05A08WE1.  Parts
of the measurements were performed at the ``Theodor Svedberg
Laboratory'' in Uppsala, Sweden. The authors wish to thank all
involved personnel 
and especially the acoustics groups of DESY Zeuthen and of Uppsala University
for their support. 



\bibliographystyle{elsarticle-num}

\end{document}